\documentclass[prd,aps,showpacs,nofootinbib,eqsecnum]{revtex4}
\usepackage{amsmath}
\usepackage{amssymb}
\usepackage{amsfonts}
\usepackage{graphicx,bm}
\usepackage{dcolumn}
\usepackage{color,amsxtra}
\usepackage{epsf}
\usepackage{enumerate}
\usepackage{hhline}
\usepackage{array}
\usepackage{tabularx}
\usepackage{subfigure}
\usepackage{fancyhdr}
\usepackage{mathrsfs}

\newcommand{\be}{\begin{equation}}
\newcommand{\ee}{\end{equation}}
\newcommand{\bea}{\begin{eqnarray}}
\newcommand{\eea}{\end{eqnarray}}
\newcommand{\beaa}{\begin{eqnarray*}}
\newcommand{\eeaa}{\end{eqnarray*}}

\newcommand{\nn}{\nonumber \\}
\newcommand{\e}{\mathrm{e}}

\def\be{\begin{equation}}
\def\ee{\end{equation}}
\def\bea{\begin{eqnarray}}
\def\eea{\end{eqnarray}}

\def\nn{\nonumber \\}
\def\e{\mathrm{e}}

\begin{document}

\title{\bf The instabilities and (anti)-evaporation of Schwarzschild-de Sitter black holes in modified gravity}

\author{
L.~Sebastiani$^{1, }$\footnote{E-mail
address: l.sebastiani@science.unitn.it},
D. Momeni$^{1, }$\footnote{E-mail address: momeni-d@enu.kz},
R. Myrzakulov$^{1, }$\footnote{E-mail address: myrzakulov@gmail.com}
and
S.~D. Odintsov$^{1,2, 3,4,}$\footnote{E-mail address: odintsov@ieec.uab.es} 
}
\affiliation{
$^1$ Department of General \& Theoretical Physics and Eurasian Center for Theoretical Physics, Eurasian National University, Astana 010008, Kazakhstan\\ 
$^2$Consejo Superior de Investigaciones Cient\'{\i}ficas, ICE/CSIC-IEEC, 
Campus UAB, Facultat de Ci\`{e}ncies, Torre C5-Parell-2a pl, E-08193
Bellaterra (Barcelona), Spain\\
$^3$Instituci\'{o} Catalana de Recerca i Estudis Avan\c{c}ats
(ICREA), Barcelona, Spain\\ 
$^4$ Tomsk State Pedagogical University, 634061,Tomsk, Russia
}


\begin{abstract}
We investigate the future evolution of Nariai black hole which is extremal 
limit of Schwarzschild-de Sitter one in modified gravity. The perturbation
equations around Nariai black hole are derived in static
and cosmological patches for general $F(R)$-gravity. The analytical and 
numerical study of several realistic $F(R)$-models shows the occurence of 
rich variety of scenarios: instabilities, celebrated Hawking evaporation 
and anti-evaporation of black holes. The realization of specific scenario 
depends on the model under consideration. It is remarkable that the 
presence of such primordial black holes at current universe may indicate 
towards the modified gravity which supports the anti-evaporation as 
preferable model. As some generalization we extend the study of Nariai 
black hole evolution to modified Gauss-Bonnet gravity. The corresponding 
perturbation equations turn out to be much more complicated than in the 
case of $F(R)$-gravity. For specific example of modified Gauss-Bonnet 
gravity we demonstrate that Nariai solution maybe stable.
\end{abstract}
\pacs{ 04.50.Kd; 04.70.Dy; 98.80.-k.}

\maketitle

\def\thesection{\Roman{section}}
\def\theequation{\Roman{section}.\arabic{equation}}

\section{Introduction}
It is well-known that 
Schwarzschild-de Sitter  solution represents the metric of black 
hole
immersed in accelerated background.
The size of black hole horizon cannot be larger than the size of de Sitter
horizon, and the properties of such black hole are not so much different
from the ones of Schwarzschild black hole in flat space. In
particulary, it is possible to associate to their horizon a
temperature~\cite{Hawking}
which is always larger than the temperature of dS-horizon. Due to the
quantum effects near to the horizon, the black holes may emit radiation
and eventually evaporate in the course of universe evolution, as Hawking predicted in 1974. The Hawking 
quantum evaporation is quite universal effect which is applied to all 
types of black holes. Schwarzschild-de Sitter black holes are primordial 
ones, they are not expected to appear at the final stage of star collapse.
Hence, such primordial black holes should evaporate and should not be 
observed at current epoch.
Nevertheless,
in the extremal limit, when the black hole horizon coincides with the
cosmological horizon~\cite{Nariai}, it has been observed by Hawking \&
Bousso~\cite{HawkingBousso} that the opposite process of anti-evaporation 
may occur.
Namely, the black hole horizon could increase, if one takes into account
the quantum corrections. The corresponding analysis has been first done in the 
so-called s-wave approximation which may not always be reliable, using the 
one-loop two-dimensional effective action. Furthermore, more consistent 
four-dimensional analysis \cite{ref.4} with account of 
full four-dimensional conformal 
anomaly also indicates to the possibility of anti-evaporation.

  A complete analysis of this phenomena has been
done in Ref.~\cite{Nie} (see also ~\cite{adventures}), where the future 
evolution of such black holes
has been investigated. This analysis has been fulfilled using again s-wave 
approximation for account of one-loop two-dimensional effective action.
A variety of situations, instabilities, anti-evaporation and evaporation, 
was discovered.

Recently, much attention has been paid to so-called modified gravity 
theories which are aimed to unify  description of early-time inflation 
and late-time acceleration of the universe (for recent review, see 
~\cite{review}). 
Gravitational action of such theory which should be effective low-energy 
limit of fundamental quantum gravity differs from the one of General 
Relativity (GR). Hence, it is quite natural to expect that some quantum gravity 
effects are encoded in modified gravity, for instance, in some versions of 
$F(R)$-gravity or string-inspired gravities. If such effective theory 
describes the 
very early universe, it is reasonable to study the properties of primordial 
black holes in terms of it. In this case, the gravitational correction to 
GR action plays the role of quantum correction. In frames of power-law 
$F(R)$-theory 
the investigation of primordial black holes evolution has been done in 
Ref.~\cite{Od}. Using static patch description it was demonstrated the 
possibility of
anti-evaporation of the so-called Nariai black hole, which is the extremal limit of Schwarzhild-de Sitter black hole.

The present work is devoted to the study of future evolution of Nariai 
black holes in viable $F(R)$ and Gauss-Bonnet modified gravities using the 
cosmological patch description. We demonstrate the rich variety of future 
evolutions: instabilities, anti-evaporation and evaporation, which depend
on the specific models under consideration. It is interesting that the 
occurence of such primordial black holes in the current epoch may serve as a
kind of test in favour of the modified gravity model which supports 
anti-evaporation processes.

The paper is organized as the following. In Section {\bf 2} the
formulation of Nariai black holes in $F(R)$-gravity is presented. The
static patch and the cosmological patch of the metric, whose geometry is
the one of $S_1$x$S_2$ are given. The comoving coordinate associated to
$S_1$-sphere is denoted with $x$.
In Section {\bf 3} the perturbations of Nariai solution are studied in the
cosmological patch of $F(R)$-gravity. Within this class of modified
gravity theories,
several models which admit the stable or unstable Schwarzschild-de Sitter
solution are introduced. In the extremal limit, the perturbations on
Nariai horizon depend on the cosmological time and on the comoving
coordinate $x$ and can be decomposed into Fourier modes, whose
coefficients are given by the Legendre polynomials.
We investigate $n=1$ mode, then generic modes and finally the
superposition of different modes for generic $F(R)$-gravities.
The Nariai horizon is generally stable when the corresponding
Schwarzschild-de Sitter solution is stable, but when the Schwarzschild-de
Sitter solution is unstable, it may be stable for some specific cases.
When the solution is unstable,  processes of evaporation or
anti-evaporation occur.
The corresponding processes are investigated numerically for number of
$F(R)$-gravities. Explicit examples of black hole (anti)-evaporation are
found analytically and numerically.
In Section {\bf 4}, the static patch of Nariai perturbations in
$F(R)$-gravity is also considered. Some different interesting behaviour
appears, namely even if the Schwarzschild-de Sitter solution is stable, it
becomes unstable in the extremal limit of the metric.
The examples of (anti)-evaporation are again discovered.
Section {\bf 5} is devoted to Nariai black holes in modified Gauss-Bonnet
$f(G)$-gravity, in the attempt to extend this study to other classes of
modified gravity. Here, in order to study the future evolution of Nariai
black holes, a complete formalism is developed. Since its results are
much more complicated with respect to the case of $F(R)$-gravity, the
analysis
is carried out for a toy model and a specific class of perturbations is
investigated.
Some outlook and Conclusions are given in Section {\bf 6}.

\section{Nariai solution in $F(R)$-gravity}

In this Section we will describe Nariai solution as the limit of Schwarzschild-de Sitter (SdS) black holes in $F(R)$-gravity.
Let us start from the action of $F(R)$-gravity in vacuum,
\begin{equation}
I=\frac{1}{2\kappa^2}\int_\mathcal{M} d^4 x \sqrt{-g}\,F(R)\,, 
\end{equation}
where $g$ is the determinant of the metric tensor $g_{\mu\nu}$,
$\mathcal M$ is the space-time manifold and $F(R)$ is a generic function of the Ricci scalar $R$.
The field equations are
\begin{equation}
F_R(R)\left(R_{\mu\nu}-\frac{1}{2}Rg_{\mu\nu}\right)=\frac{1}{2}g_{\mu\nu}[F(R)-RF_R(R)]
+(\nabla_{\mu}\nabla_{\nu}-g_{\mu\nu}\Box)F_R(R)\,.
\label{fieldequation}
\end{equation}
Here, $F_R=\partial F(R)/\partial R$. 
Furthermore, ${\nabla}_{\mu}$ is the covariant derivative
operator associated with $g_{\mu \nu}$, and
$\Box\equiv g^{\mu\nu}\nabla_{\mu}\nabla_{\nu}$ is the
covariant d'Alembertian.
The SdS solution reads
\begin{equation}
ds^2=-B(r)dt^2+\frac{dr^2}{B(r)}+r^2d\Omega^2\,,
\end{equation}
where $d\Omega^2=\sin\phi^2 d\theta^2+d\phi^2$ is the metric of the $S_2$ sphere and $B(r)$ is a function of the radial coordinate $r$,
\begin{equation}
B(r)=1-\frac{2 M}{r}-\frac{\Lambda}{3}r^2\,.\label{solution}
\end{equation}
Here, $M>0$ is a mass parameter and $\Lambda$ is a cosmological constant related with the curvature as $R_0=4\Lambda$ (we assume there is a constant curvature solution $R_0$).
The SdS metric represents the static, spherically symmetric solution of GR with cosmological constant $\Lambda$ (i.e. $F(R)=R-2\Lambda$). Many $F(R)$-gravity models admit this kind of solution under the condition~\cite{CognolaJCAP}
\begin{equation}
2F(R_0)=R_0 F_R(R_0)\,,\label{dScondition}
\end{equation}
which follows from Eq. (\ref{fieldequation}) in the case of constant curvature. It is worth to mention that in $F(R)$-gravity the cosmological constant $\Lambda$ is usually fixed by the model, except for the case of $R^2$-gravity, which admits the SdS solution with free $\Lambda$. This is a special case, since the action does not contain a fundamental length scale
which depends on $\kappa^2$ and the lenght scale is provided by the solution ( $[\Lambda]=[r^{-2}]$). 
Note that black hole solutions in $F(R)$-gravity have been studied in the number of works, see for example Refs.~\cite{SSSsolutions,S,energy,trtrtr}.

Let us return to SdS solution in (\ref{solution}).
For $0<M<1/(3\sqrt{\Lambda})$, we have two positive roots $r_+$ and $r_{++}$ of $B(r)=0$, which correspond to the black hole horizon (in such a case, $B'(r_+)>0$) and to the cosmological horizon ($B'(r_{++})<0$). When $M\rightarrow 0^+$, we recover the pure de Sitter solution, and when $M\rightarrow 1/(3\sqrt{\Lambda})$, the black hole horizon approaches the cosmological horizon and the coordinate $r$ becomes degenerate and meaningless. For this reason, in the extremal limit one introduces the coordinates $\psi$ and $\chi$ related with $t$ and $r$ as
\begin{equation}
t=\frac{1}{\epsilon\sqrt{\Lambda}}\psi\,,\quad r=\frac{1}{\sqrt{\Lambda}}\left[1-\epsilon \cos\chi-\frac{1}{6\epsilon^2}\right]\,, \label{6}
\end{equation}
where, following Ref.~\cite{GP}, we write $9M^2\Lambda=1-3\epsilon^2$ with $0\leq\epsilon\ll 1$. The degenerate case is given by $\epsilon\rightarrow 0$. 
Now black hole and cosmological horizons correspond to the cases $\chi=0\,,\pi$.
The following metric is derived:
\begin{equation}
ds^2 =-\frac{1}{\Lambda}\left(1+\frac{2}{3}\epsilon\cos\chi\right)\sin^2\chi d\psi^2+\frac{1}{\Lambda}\left(1-\frac{2}{3}\epsilon\cos\chi\right)d\chi^2+\frac{1}{\Lambda}(1-2\epsilon\cos\chi)d\Omega^2\,.\label{Haw}
\end{equation}
This is the SdS solution (nearly the maximal case). The topology is given by $S_1\times S_2$.
The degenerate case corresponds to the
Nariai space-time~\cite{Nariai}, namely
\be
\label{Nr1}
ds^2 = \frac{1}{\Lambda} \left(-\sin^2\chi d\psi^2+d\chi^2\right)
+ \frac{1}{\Lambda}d\Omega^2\,.
\ee
Note that in this case the two spheres of the horizons have the same radius $r_H=1/\sqrt{\Lambda}$. The interesting point is that, since the radial coordinate $r$ in (\ref{6}) is degenerate, one obtains $B(r)=0$ in the extremal limit of SdS solution, but in this coordinate system the singularity is absorbed by the transformation. The metric is regular and describes the geometry of Nariai solution.

In order to arrive to the cosmological patch of Nariai solution, we firstly reintroduce the singularity in (\ref{Haw}) through $\chi=-\arcsin z$, such that
\begin{equation}
ds^2=-\frac{1}{\Lambda}\left(1-z^2\right)d\psi^2+\frac{dz^2}{\Lambda(1-z^2)}+\frac{1}{\Lambda}d\Omega^2\,,
\end{equation}
which is singular for $z=\pm 1$. Thus, we write the time coordinate $t$ and the comoving coordinate $x$ as
\begin{eqnarray}
t=\psi+\frac{1}{2}\log(1-z^2)\,,\quad
x=\frac{z}{(1-z)^{1/2}}\mathrm{e}^{\pm t}\,.
\end{eqnarray}
In the global coordinates, the Nariai metric finally reads
\begin{equation}
ds^2= \frac{1}{\Lambda}(-dt^2+\mathrm{e}^{\pm 2t}dx^2)
+ \frac{1}{\Lambda}d\Omega^2\,.\nonumber
\end{equation}
More in general, it is possible to find that
\begin{equation}
ds^2= \frac{1}{\Lambda}(-dt^2+\cosh^2t\, dx^2)
+ \frac{1}{\Lambda}d\Omega^2\,,\label{cosmpatch}
\end{equation}
is the most general form of Nariai metric in the cosmological patch with $R_0=4\Lambda$. 
The geometry remains
the one of $S_1\times S_2$. The $S_2$ sphere is constant with radius 
$1/\sqrt{\Lambda}$, while the radius of $S_1$ sphere expands exponentially as $\mathrm{e}^{2t}/(\Lambda)$. 
Note that based on classical solution of this Section one can develop
the
picture of quantum creation and nucleation of Nariai black holes in $F(R)$
gravity in the close analogy with Ref.~\cite{Nie}.
Of course, the corresponding problem is more complicated in modified
gravity
and requests a separate study.

\section{Nariai solution in the cosmological patch of $F(R)$-gravity}

In this Section, we will analyze the cosmological patch of Nariai solution. Perturbations on the horizon are found and studied in different $F(R)$-gravity models, where future evolution of Nariai black holes is investigated.

The cosmological patch of Nariai solution (\ref{cosmpatch}) can be written in the following equivalent form,
\begin{equation}
ds^2=-\frac{1}{\Lambda\cos^2 \tau}\left(-d\tau^2+dx^2\right)+\frac{1}{\Lambda}d\Omega^2\,,\label{Nariai}
\end{equation}
where 
\begin{equation}
\tau=\arccos\left[\cosh t\right]^{-1}\,,\label{transf}
\end{equation}
such that it is easy to see that $-\pi/2<\tau<\pi/2$ corresponds to $-\infty<t<+\infty$.
This metric describes the further evolution of the Nariai solution.

We now consider the perturbations from the Nariai
space-time.
One assumes the following metric Ansatz:
\be
\label{Nr3}
ds^2 = \e^{2\rho\left(x,\tau\right)} \left( - d\tau^2 + dx^2 \right)
+ \e^{-2 \varphi\left(x,\tau\right)} d\Omega^2\,.
\ee
The Ricci scalar reads
\begin{equation}
R=\left(2\ddot\rho-2\rho''-4\ddot\phi+4\phi''+6\dot\phi^2-6\phi'^2\right)\mathrm{e}^{-2\rho}
+2\mathrm{e}^{2\phi}\,.
\end{equation}
The dot denotes the derivative with respect to $\tau$ and the prime denotes the derivative with respect to $x$.
The $(0,0)$, $(1,1)$, $(0,1)$
$\left(=\left(1,0\right) \right)$, and
$(2,2)$ $\left(=\left(3,3\right)\right)$
components of (\ref{fieldequation}) take the following forms:
\begin{align}
\label{Nr5}
0 = & - \frac{\e^{2\rho}}{2} F(R) - \left( - \ddot\rho + 2 \ddot\varphi
+ \rho'' - 2 {\dot\varphi}^2 - 2 \rho' \varphi' - 2 \dot\rho \dot\varphi
\right) F_R(R)
+ \frac{\partial^2 F_R(R)}{\partial\tau^2} - \dot\rho \frac{\partial
F_R(R)}{\partial\tau} - \rho' \frac{\partial F_R(R)}{\partial x} \nn
& + \e^{2\varphi} \left\{ - \frac{\partial}{\partial\tau}
\left( \e^{-2\varphi} \frac{\partial F_R(R)}{\partial\tau}\right)
+ \frac{\partial}{\partial x}
\left( \e^{-2\varphi} \frac{\partial F_R(R)}{\partial x}\right)\right\}
\, ,\nn
0 = & \frac{\e^{2\rho}}{2} F(R) - \left( \ddot\rho + 2 \varphi''
 - \rho'' - 2 {\varphi'}^2 - 2 \rho' \varphi' - 2 \dot\rho \dot\varphi \right)
F_R(R)
+ \frac{\partial^2 F_R(R)}{\partial x^2} - \dot\rho \frac{\partial
F_R(R)}{\partial\tau} - \rho' \frac{\partial F_R(R)}{\partial x} \nn
& - \e^{2\varphi} \left\{ - \frac{\partial}{\partial\tau}
\left( \e^{-2\varphi} \frac{\partial F_R(R)}{\partial\tau}\right)
+ \frac{\partial}{\partial x}
\left( \e^{-2\varphi} \frac{\partial F_R(R)}{\partial x}\right)\right\}
\, ,\nn
0 = & - \left( 2 {\dot\varphi}' - 2 \varphi' \dot\varphi - 2\rho' \dot\varphi
 -2 \dot\rho \varphi' \right) F_R(R)
+ \frac{\partial^2 F_R(R)}{\partial\tau \partial x} - \dot\rho \frac{\partial
F_R(R)}{\partial x} - \rho' \frac{\partial F_R(R)}{\partial\tau} \, ,\nn
0 = & \frac{\e^{- 2\varphi}}{2} F(R) - \e^{-2 \left(\rho+\varphi\right)}
\left( - \ddot\varphi + \varphi'' - 2 {\varphi'}^2 + 2 {\dot\varphi}^2 \right)
F_R(R) - F_R(R) + \e^{-\left(\rho+\varphi\right)} \left(
\dot\varphi \frac{\partial F_R(R)}{\partial\tau}
 - \varphi' \frac{\partial F_R(R)}{\partial x} \right)\nn
& - \e^{- 2\rho} \left\{ - \frac{\partial}{\partial\tau}
\left( \e^{-2\varphi} \frac{\partial F_R(R)}{\partial\tau}\right)
+ \frac{\partial}{\partial x}
\left( \e^{-2\varphi} \frac{\partial F_R(R)}{\partial x}\right)\right\}\, .
\end{align}
The perturbations of the Nariai space-time (\ref{Nariai}) could be written in terms of $\delta\rho(\tau,x)$ and $\delta\varphi(\tau,x)$,
\be
\label{Nr6}
\rho = - \ln \left[ \sqrt{\Lambda} \cos \tau \right] + \delta\rho\, ,\quad
\varphi = \ln \sqrt{\Lambda} + \delta \varphi\,.
\ee
Then one gets
\be
\delta R = 4 \Lambda \left( - \delta\rho + \delta\varphi \right)
+ \Lambda \cos^2 \tau \left( 2 \delta \ddot\rho - 2 \delta \rho''
 - 4 \delta \ddot\varphi + 4\delta \varphi'' \right) \, .\label{varR}
\ee
Let us assume that our $F(R)$-gravity admits the Nariai solution for $R=R_0$,
such that condition (\ref{dScondition}) is satisfied.
At the first order,
the perturbed equations from (\ref{Nr5}) are:
\begin{align}
\label{Nr8}
0 = & \frac{- F_R\left( R_0 \right) + 2 \Lambda F_{RR}\left( R_0 \right)}{2
\Lambda \cos^2 \tau} \delta R
 - \frac{F \left( R_0 \right)}{\Lambda \cos^2 \tau}\delta\rho
 - F_R \left( R_0 \right) \left( - \delta \ddot\rho + 2 \delta \ddot\varphi
+ \delta \rho'' - 2 \tan \tau \delta\dot\varphi \right) \nn
& - \tan \tau F_{RR} \left( R_0 \right) \delta \dot R + F_{RR} \left( R_0 \right)
\delta R''\, ,\nn
0 =& -  \frac{- F_R\left( R_0 \right) + 2 \Lambda F_{RR}\left( R_0 \right)}{2
\Lambda \cos^2 \tau} \delta R
+ \frac{F \left( R_0 \right)}{\Lambda \cos^2 \tau}\delta\rho
 - F_R \left( R_0 \right) \left( \delta \ddot\rho + 2 \delta \varphi''
 - \delta \rho''
- 2 \tan \tau \delta\dot\varphi \right)\nn 
&+ F_{RR} \left( R_0 \right) \delta \ddot R-\tan\tau F_{RR}(R_0)\delta \dot R
\, ,\nn
0 = & - 2 F_R(R_0) \left( \delta{\dot\varphi}' - \tan \tau \delta \varphi' \right)
+ F_{RR} \left( R_0 \right) \left(
\delta {\dot R}' - \tan \tau \delta  R'\right)\, ,\nn
0 = & - \frac{- F_R\left( R_0 \right) + 2 \Lambda F_{RR} \left( R_0 \right)}{2
\Lambda } \delta R - \frac{F \left( R_0 \right)}{\Lambda} \delta\varphi
 - \cos^2 \tau F_R \left( R_0 \right) \left( - \delta \ddot\varphi + \delta
\varphi'' \right)\nn
& - \cos^2 \tau F_{RR} \left( R_0 \right) \left( - \delta \ddot R + \delta R''
\right) \, .
\end{align}
The third equation can be integrated to give
\be
\delta R=2\frac{F_R(R_0)}{F_{RR}(R_0)}\delta\varphi+\frac{C_x(x)}{\cos\tau}+C_\tau(\tau)
\, .\label{deltaR}
\ee
Here, $C_x(x)$ and $C_\tau(\tau)$ are arbitrary functions of $x$ and $\tau$,
respectively. In what follows, we put $C_x(x)=C_\tau(\tau)=0$. By inserting this result in the first two equations and by using Eq. (\ref{varR}) we find that they are trivially satisfied. Finally, the fourth equation leads to
\begin{equation}
\frac{1}{\alpha\cos^2\tau}\left[2(2\alpha-1)\delta\varphi\right]-3\delta\ddot\varphi+3\delta\varphi''=0\,,
\label{principe}
\end{equation}
where it is defined
\begin{equation}
\alpha=\frac{2\Lambda F_{RR}(R_0)}{F'(R_0)}\,.\label{alpha}
\end{equation}
Equation (\ref{principe}) can be used to study the evolution of $\varphi(\tau,x)$. In principle, one may insert the result in (\ref{deltaR}) in order to obtain $\rho(\tau,x)$. However, the radius of the Nariai black hole depends on $\varphi(\tau,x)$ only, so that we will limit our analysis to Eq. (\ref{principe}). 
As it was observed in Ref.~\cite{Nie}, the position of the horizon moves on the one-sphere $S_1$.
More specifically, it is located in the correspondence of
\begin{eqnarray}
\nabla\delta\varphi\cdot\nabla\delta\varphi=0\,,\label{horizon}
\end{eqnarray}
namely it is required that the (flat) gradient of the two-sphere size is null. For a black hole located at $x=x_0$, the horizon is defined as
\begin{equation}
r_0(\tau)^{-2}=\mathrm{e}^{2\varphi(\tau,x_0)}=\frac{1+\delta\varphi(x_0,\tau)}{\Lambda}\,.
\label{relation}
\end{equation}
Therefore, evaporation/anti-evaporation correspond to increasing/decreasing values of $\delta\varphi(\tau)$ on the horizon.

\subsection{Schwarzschild-de Sitter black holes in viable $F(R)$-gravity}

Several versions of viable
modified
$F(R)$-gravity have been proposed in literature in order to
reproduce the current acceleration of the universe as well as inflationary scenario (for recent review of unified description of inflation with dark energy in modified gravity, see Refs.~\cite{review}).
The $F(R)$-models for current dark energy epoch
must satisfy a list of viability conditions. 
In particular, since the results of General Relativity were first confirmed by local tests
at the level of the Solar System, such a kind of modified gravity has to
admit a
static spherically symmetric solution, typically the Schwarzschild or the Schwarzschild-de Sitter solution. In order to avoid significant corrections to the Newton law (fifth force), this solution must be stable. The stability of SdS solution (or dS-solution) is provided by the well-know condition
\begin{equation}
0<\alpha<\frac{1}{2}\,,\label{stabBH}
\end{equation}
which directly follows from the time-perturbation of the trace of the field equations in vacuum~\cite{qualcosa}.
From Eq. (\ref{principe}), if we restrict to the case $\delta\varphi(\tau,x)=\delta\varphi(\tau)$, by reintroducing the cosmological time $t$ via (\ref{transf}), one obtains
\begin{equation}
\frac{d^2\delta\varphi}{d t^2}+\tanh t\frac{d\delta\varphi}{d t}-m^2\delta\varphi=0\,,\quad m^2=\frac{2(2\alpha-1)}{3\alpha}\,,
\end{equation}
where $\alpha$ depends on the model as in (\ref{alpha}). 
Since we are interested in the future evolution of Nariai horizon, we can take $t\gg 0$, such that $\tanh t\simeq 1$. In this case the solution reads
\begin{equation}
\delta\varphi(t)=\varphi_0\mathrm{e}^{\left(\frac{-1\pm\sqrt{1+4m^2}}{2}\right)\,t}\,,
\end{equation}
where $\varphi_0$ is a generic constant. As a consequence, we recover the stability condition (\ref{stabBH}) of SdS solution: this condition always is valid when one considers time-perturbations of a metric with constant curvature. In particular, when $0<\alpha<8/19$, an imaginary part appears and the solution oscillates around the horizon and does not diverge.

In the present work, we will analyze the Nariai solution by taking into account 
the $x$-dependence of perturbation starting from Eq.~(\ref{principe}). 
Since it could be interesting to investigate the behaviour of Nariai solution in some specific $F(R)$-gravity, we conclude this Subsection by introducing some specific models.
At first, we present a simple class of $F(R)$-gravity for the dark energy
which describes the stable SdS solution.
In these models, 
a correction term 
to the Hilbert-Einstein action is added 
as $F(R)=R+f(R)$, being $f(R)$ a generic function of the Ricci scalar, and
the dark energy epoch is produced in a simple way:
a vanishing cosmological constant in the flat limit 
of $R=0$ is incorporated, and a suitable, constant asymptotic 
behavior for large values of $R$ is exhibited and mimics an effective cosmological constant.
In Refs.~\cite{Starobinsky:2007hu,HuSaw,Cognola:2007zu,Linder:2009jz,Battye} several versions of this kind of (viable)
modified
$F(R)$-gravity models have been proposed. Here, we present two examples. 

The first one is the Hu-Sawicki model~\cite{HuSaw}, namely
\begin{equation}
F(R)=R-\frac{\tilde{m}^{2}c_{1}(R/\tilde{m}^{2})^{n}}{c_{2}(R/\tilde{m}^{2})^{n}+1}=R-\frac{\tilde{m}^{2}c_{1}}{c_{2}}+\frac{\tilde{m}^{2}c_{1}/c_{2}}{c_{2}(R/\tilde{m}^{2})^{n}+1}\,,
\label{HuSawModel}
\end{equation}
where $\tilde{m}^{2}$ is a mass scale, $c_{1}$ and $c_{2}$ are positive parameters, and $n$ is a natural positive number. 
In this model 
$\tilde{m}^{2} c_{1}/c_{2}=2\Lambda_{\text{eff}}$ is an effective
cosmological constant and thus 
the $\Lambda$CDM model can be easily mimicked for large curvature. 
We reparametrize the model by 
putting $\tilde{m}^{2}c_{1}/c_{2}=2\Lambda_{\text{eff}}$ and 
$(c_2)^{1/n}\,\tilde{m}^2=\Lambda_{\text{eff}}$ with $n=4$, so that we obtain 
\begin{equation}
F(R)=R-2\Lambda_{\text{eff}}\left\{1-\frac{1}{\left[R/\Lambda_{\text{eff}}\right]^{4}+1}\right\}\,.
\label{model2}
\end{equation}
With this parametrization the Hu-Sawicki model satisfies all the cosmological constraints.
Moreover, 
in Refs.~\cite{Cognola:2007zu, Linder:2009jz, one} 
another simple exponential model has been constructed. A viable version is given by 
\begin{equation}
F(R)=R-2\Lambda_{\text{eff}}\left[1-\mathrm{e}^{-R/\Lambda_{\text{eff}}}\right]\,.
\label{model}
\end{equation}
Also in this model, for flat space the Minkowski solution 
is recovered, and at large curvatures 
the $\Lambda$CDM model is realized. 
Both of these models satisfy the cosmological and local gravity constraints, but 
the approaching to $\Lambda$CDM model is realized in two different ways, namely
via a power function of $R$ (the first one) and 
via an exponential function of it (the second one). 
These models have SdS solution (and therefore, the Nariai solution) for $R_0= 4\Lambda$ and $\Lambda\simeq\Lambda_{\text{eff}}$. A direct numerical evaluation from (\ref{dScondition}) gives us $R_0=3.95\Lambda_{\text{eff}}$ ($\Lambda=0.99\Lambda_{\text{eff}}$) for Hu-Sawicki model and $R_0=3.74\Lambda_{\text{eff}}$ ($\Lambda=0.94\Lambda_{\text{eff}}$)  for the exponential one. The associated values of $\alpha$ (\ref{alpha}) are $\alpha=0.02$ for Hu-Sawiki model and $\alpha=0.09$ for exponential gravity. 

Moreover, in the attempt to explain the phenomenology of the inflation, the power-law behaviour of Ricci scalar is often used. This kind of terms also may protect the theory against future singularities. If the inflation is described by the de Sitter solution, it has to be unstable. An example is given by $F(R)$-gravity in the form
\begin{equation}
F(R)=R+\gamma R^m\,,\label{inflmodel}
\end{equation}
where $\gamma$ is a constant dimensional parameter and $m$ is a positive number. The dS solution occurs at $R=R_0$ which solves Eq.~(\ref{dScondition}), namely
\begin{equation}
 R_{0}=\left(\frac{1}{\gamma(m-2)}\right)^{\frac{1}{m-1}}\,,\quad m\neq 2\,,\nonumber
\end{equation}
and
\begin{equation}
\Lambda=\frac{1}{4}\left(\frac{1}{\gamma(m-2)}\right)^{\frac{1}{m-1}}\,,\quad\alpha=\frac{m-1}{2\left(1+\frac{m-2}{m}\right)}\,.
\nonumber
\end{equation}
It is easy to verify that this solution violates the stability condition (\ref{stabBH}) when $m>2$,
$\gamma>0$. In the following, some examples of this kind of models producing unstable de Sitter (and the corresponding SdS) solution will be considered, since it could be interesting to know the evolution of Nariai black holes in primordial universe described by $F(R)$-gravity. 

\subsection{Horizon perturbations}

The equation (\ref{principe}) 
belongs to the class of Hamilton Jacobi equations.
Following Ref. \cite{Nie}, we decompose the two-sphere radius of Nariai solution into Fourier modes on the $S_1$ sphere, namely
\begin{equation}
\delta\varphi(x,t)=\epsilon\sum_{n=1}^{+\infty}\left(A_n(\tau)\cos[n x]+B_n(\tau)\sin[n x]\right)\,,
\quad1\gg \epsilon>0\,.
\end{equation}
Here, $\epsilon$ is assumed to be positive and small. By means of 
this expression,
we obtain the following equations for $A_n(\tau)$ and $B_{n}(\tau)$ from Eq. (\ref{principe}):
\begin{eqnarray}
\left\{
\begin{array}{l}
3\ddot A_n(\tau)\alpha\cos^2\tau-A_n(\tau)(4\alpha-2-3n^2\alpha\cos^2\tau)=0\\\\
3\ddot B_n(\tau)\alpha\cos^2\tau-B_n(\tau)(4\alpha-2-3n^2\alpha\cos^2\tau)=0\,.
\end{array}\right.
\end{eqnarray}
We rewrite this system as
 \begin{eqnarray} \frac{d^2C_{n}(t)}{d\tau^2}+\Big(\frac{3n^2\alpha\cos(\tau)^2+2(1-2\alpha)}{3\alpha\cos(\tau)^2}\Big)C_{n}(\tau)=0\label{eqc}\,,
 \end{eqnarray}
where $C_{n}(\tau)=\{A_{n}(\tau),B_{n}(\tau)\}$. By putting $\sin(\tau)=\xi$, such that
$0<\xi<1$, one derives the associated Legendre equation
 \begin{eqnarray}
 (1-\xi^2)\frac{d^2 C_{n}(\xi)}{d^2\xi}-2\xi \frac{d C_{n}(\xi)}{d\xi}+\Big[\nu(1+\nu)-\frac{\mu^2}{1-\xi^2}\Big]C_{n}(\xi)=0\,.\label{xieq}
 \end{eqnarray}
Here, 
 \begin{equation}
 \mu=\sqrt{\frac{2(2\alpha-1)}{3\alpha}}\,,\quad\nu=-\frac{1}{2}\pm\sqrt{n^2+\frac{1}{4}}\,.
 \end{equation}
In this formalism, $\mu$ depends on the $F(R)$-gravity model which admits the SdS (and Nariai) solution for $R_0=4\Lambda$ and $\nu$ on the perturbation mode $n$. Moreover, in Eq. (\ref{xieq}) $\mu$ and $\nu$ may get real or complex values.
Since $0<\xi<1$,  and generally $\mu$ is not an integer number, the solutions of this equation are the Legendre polynomials. The Legendre polynomials which are regular on the boundary coordinate $\xi=0$ (it corresponds to the cosmological time $t=0$) are
\begin{eqnarray}
P_{\nu}^{\mu}(\xi)=2^{\mu}\pi^{1/2}(\xi^2-1)^{-\mu/2}\Big[\frac{F(-\frac{\nu+\mu}{2},\frac{1+\nu-\mu}{2},\frac{1}{2};\xi^2)}{\Gamma(\frac{1-\nu-\mu}{2})\Gamma(1+\frac{\nu-\mu}{2})}-2\xi\frac{F(\frac{1-\nu-\mu}{2},1+\frac{\nu-\mu}{2},\frac{3}{2};\xi^2)}{\Gamma(\frac{1+\nu-\mu}{2})\Gamma(-\frac{\nu+\mu}{2})}\Big],\ \ |\xi^2|<1\,,
\end{eqnarray}
where $F(a,b,c;z)$ represents the hypergeometric series and $\Gamma(z)$ is the Euler function. This formula is valid in our range $-1<\xi<1$.
Now, the metric perturbation can be written as 
 \begin{eqnarray}
 \delta\varphi(x,t)=\epsilon\sum_{n=1}^{\infty}P_{\nu}^{\mu}(\xi)\Big[a_n\cos(nx)+b_n\sin(nx)\Big]\,,\label{finalpert}
 \end{eqnarray} 
where the unknown coefficients $\{a_n,b_n\}$ can in principle be obtained by using the initial  boundary conditions at $t=0$ ($\xi=0$). To be more precise, we may think the previous expression in terms of the orthonormal basis functions $|\cos(nx)>,|\sin(nx)>$, namely
\begin{eqnarray}
| \delta\varphi(x,t)>=\epsilon\sum_{n=1}^{\infty}P_{\nu}^{\mu}(\xi)\Big[a_n|\cos(nx)>+b_n|\sin(nx)>\Big]\,.
\end{eqnarray}
The unique evolutionary scheme of the wave packet function $| \delta\varphi(x,t)>$ is fixed by Cauchy's boundary conditions at $| \delta\varphi(x,0)>$ and $|\frac{\partial \delta\varphi(x,0)}{\partial \xi}>$.
As a consequence, by multiplying  $| \delta\varphi(x,0)>$ with $<\cos nx|$ and 
$|\frac{\partial \delta\varphi(x,0)}{\partial \xi}>$ with $<\sin nx|$, we derive
the coefficents $\{a_n,b_n\}$ in the following integral forms,
\begin{eqnarray} 
a_n=\frac{1}{\pi \epsilon P_{\nu}^{\mu}(0)}\int_{0}^{\pi}dx \delta\varphi(x,0)\cos(nx)\,,\quad
b_n=\frac{1}{\pi \epsilon (d P_{\nu}^{\mu}(0)/d\xi)}\int_{0}^{\pi}dx \frac{\partial \delta\varphi(x,0)}{\partial \xi}\sin(nx)\,.\label{bum}
 \end{eqnarray}
Here,
\begin{eqnarray}
P_{\nu}^{\mu}(0)=2^{\mu}\pi^{-1/2}\frac{\cos[\frac{\pi(\mu+\nu)}{2}]\Gamma(\frac{1+\nu+\mu}{2})}{\Gamma(\frac{1+\nu-\mu}{2})}.
\end{eqnarray}
The horizon perturbation, namely the perturbation on the position of Nariai horizon $\delta\phi(x_0,\tau)$ of expression (\ref{relation}), can be derived from (\ref{finalpert}) by imposing condition (\ref{horizon}). 

\subsubsection{Mode $n=1$}

\begin{figure}
\begin{center}
\includegraphics[angle=0, width=0.40\textwidth]{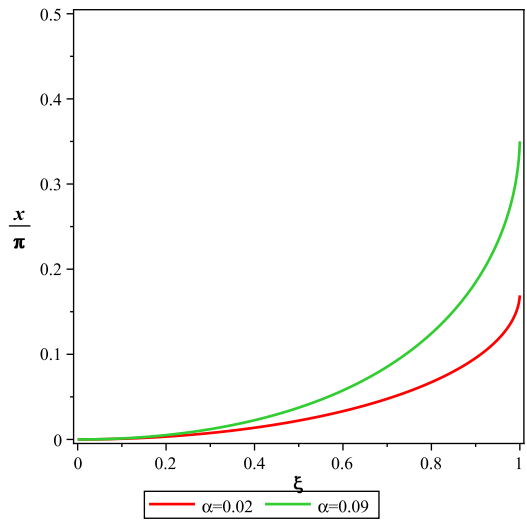}
\end{center}
\caption{Evolution of horizon on the $S_1$ coordinate $0<x<\pi/2$ for the perturbation mode $n=1$ for Hu-Sawicki model ($\alpha=0.02$) and exponential gravity ($\alpha=0.09$).\label{Fig1}}
\end{figure}

In this Subsection we restrict our analysis to the first mode perturbation $n=1$ 
in Eq.~(\ref{finalpert}). It means
\begin{equation}
\nu=-\frac{1}{2}\pm\sqrt{\frac{5}{4}}\,.
\end{equation}
In what follows, we will consider the case of the plus sign. 
Morever, since in fact $a_1$ encodes the initial shape and $b_1$ the initial velocity of perturbation,
by following Ref.~\cite{Nie} we write this paramters as
\begin{equation}
a_1=\sin\theta\,,\quad b_1=\cos\theta\,,\label{ab}
\end{equation} 
being $\theta$ a fixed angular coordinate. Changing $\theta$ one may represent different initial physical configurations.
By using condition (\ref{horizon}), we obtain
\begin{equation}
\tan x=\pm\frac{b_1-\mathcal A(\mu,\nu;\xi)a_1}{a_1+\mathcal A(\mu,\nu;\xi)b_1}\,,
\label{43}
\end{equation}
where
\begin{equation}
\mathcal A(\mu,\nu;\xi)=\frac{(d P^{\mu}_{\nu}(\xi)/d\xi)}{P^{\mu}_{\nu}(\xi)}\sqrt{1-\xi^2}\,.
\label{44}
\end{equation}
Note that, since
\begin{equation}
\frac{d P^\mu_\nu(\xi)/d\xi}{P^\mu_\nu(\xi)}=\frac{(\nu+1)\xi P^\mu_\nu(\xi)-(\nu-\mu+1)P^\mu_{\nu+1}(\xi)}{(1-\xi^2)P^\mu_\nu(\xi)}\,,
\end{equation}
in general $\mathcal A(\mu,\nu;\xi)$ is real even if $\mu$ and therefore the Legendre polynomial are imaginary.
In Fig.~\ref{Fig1} we show the location of the horizon on the $S_1$ sphere in the cases of Hu-Sawicki model ($\alpha=0.02$; $\mu=5.66i$) and in the case of exponential gravity ($\alpha=0.09$; $\mu=2.46i$) in the range $0<x<\pi/2$. We have set 
\begin{equation}
\frac{b}{a}=\mathcal A(\mu,\nu;0)\,,
\end{equation}
in order to locate the horizon in $x=0$ at the time $\tau(\equiv t)=0$.

The perturbation on the horizon reads (for the positive value of the tangent)
\begin{eqnarray}\label{pert}
\delta\varphi(\xi)=\epsilon\,P^{\mu}_{\nu}(\xi)\left\{\frac{a_1}{\sqrt{1+\left(\frac{b_1-\mathcal A(\mu,\nu;\xi)a_1}{a_1+\mathcal A(\mu,\nu;\xi)b_1}\right)^2}}+
\frac{b_1}{\sqrt{1+\left(\frac{a_1+\mathcal A(\mu,\nu;\xi)a_1}{b_1-\mathcal A(\mu,\nu;\xi)b_1}\right)^2}}
\right\}\,.
\end{eqnarray} 
In order to study the evolution of Nariai solution, one must investigate the behaviour of the Legendre polynomial $P_\mu^\nu(\xi)$ near $\xi=1$, since
from the definition of $\tau$ (\ref{transf}), and therefore of $\xi$, we can see an infinite amount of time, i.e. the future evolution of the solution, is contained in this limit.
In this case $\nu$ is real, and we have two different solutions for $\mu$ positive and real, namely $\alpha>1/2$ or $0>\alpha$, and $\mu$ imaginary, it means $0<\alpha<1/2$, which correspond to unstable and stable SdS solution, respectively. When $\alpha$ is real, we have 
\begin{eqnarray}
P^\mu_\nu(\xi)\simeq (1-\xi)^{-\frac{\mu}{2}}\left[\frac{2^{\mu/2}}{\Gamma(1-\mu)}-\frac{2^{\mu/2}(\mu-\mu^2+2\nu(1+\nu))}{4\Gamma(2-\mu)}(1-\xi)+\mathcal{O}(1-\xi)^2\right]\,,&&\nonumber\\\mu\in R^+\,,(1-\xi)\ll1\,.\label{mureal}
\end{eqnarray}
As a consequence, when $\mu$ is real, the Legendre function and therefore the perturbation asymptotically diverges and grows up. This effect corresponds to anti-evaporation. 
However, a different behaviour is obtained if $\mu$ is an integer number: in this case the Gamma function posses the poles and
the first terms of the expansion tend to zero. As a consequence, the analysis has to be done by using the lowest terms and the solution is generally stable. An example is given by $\mu=1$ (it corresponds to $\alpha=2$), for which we derive at the first order,
\begin{equation}
P^\mu_\nu(\xi)\simeq -\frac{\sqrt{2}}{2}(1-\xi)^{1/2}\,,\quad\mu=1\,,(1-\xi)\ll 1\,,\label{according}
\end{equation}
where the fact that $2\nu(1+\nu)=1$ is used. In this special case, the solution is stable.
Some important remarks are in order. We have found that, in general, when $\alpha>1/2$ or $0>\alpha$ the Nariai solution is unstable and we have anti-evaporation at $t\rightarrow+\infty$. Of course, some transient effects of evaporation for small value of $t$ are not excluded. Moreover, we have 
choosen as the initial condition $\epsilon>0$ (but it does not necessarly mean that the initial perturbation is positive). If $\epsilon$ is negative, we obtain the opposite result. Stability/unstability of the solution does not depend on the sign of $\epsilon$, but if the solution is unstable the final evolution of Nariai solution depends on it. In particular, if $\epsilon$ is negative, we have an evaporation process.

When $\mu$ is imaginary ($0<\alpha<1/2$), if $\mu=i|\mu|$ where $|z|$ is the norm of $z$, we get in the limit $\xi\rightarrow 1^-$,
\begin{eqnarray}
P^{i|\mu|}_\nu(\xi)\simeq (1-\xi)^{-\frac{i|\mu|}{2}}\left[\frac{2^{\frac{i|\mu|}{2}}}{\Gamma(1-i|\mu|)}-\frac{2^{\frac{i|\mu|}{2}}(1-\xi)}{4\Gamma(2-i|\mu|)}(|\mu|(i+|\mu|)+2\nu(\nu+1))+\mathcal{O}(1-\xi)^2)
\right]\,,&&\nonumber\\
\mu\in C\,,(1-\xi)\ll 1\,.
\label{50}
\end{eqnarray} 
The real part of this expression reads
\begin{equation}
\mathcal{R}(P_\nu^{i|\mu|}(\xi))\simeq\frac{1}{|\Gamma(1-i|\mu|)|}\cos\left[\frac{|\mu|}{2}\log\left[\frac{2}{1-\xi}\right]-\phi\right]\,,
\end{equation}
where
\begin{equation}
\Gamma(1-i|\mu|)=|\Gamma(1-i|\mu|)|\mathrm{Exp}\left[i\phi\right]\,,\quad\phi=\mathrm{arg}[\gamma(1-i|\mu|)]\,.
\end{equation}
As the result, one may conclude that for the mode $n=1$, Nariai solution of $F(R)$-gravity with $0<\alpha<1/2$ is stable, oscillating around the horizon and passing from evaporation to anti-evaporation region for an infinite number of times. We will show in the next Subsection that some instabilities may appear when we will consider the static patch of Nariai solution.

\begin{figure}
\begin{center}
\includegraphics[angle=0, width=0.40\textwidth]{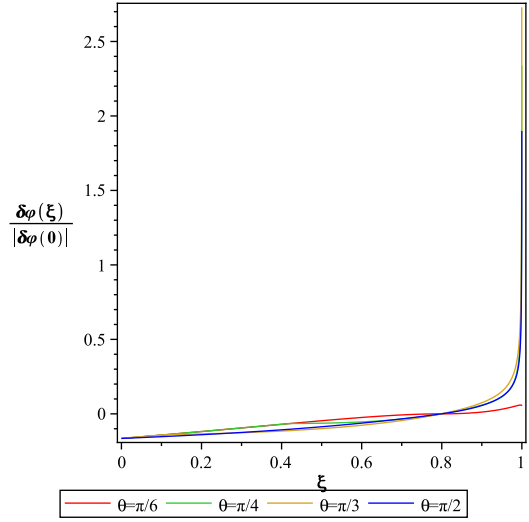}
\end{center}
\caption{\label{Fig2a} Evolution of Nariai horizon perturbation for $n=1$ and different choices of $\theta$ in the model $F(R)=R+\gamma R^5$. The horizon grows up and finally diverges. Anti-evaporation process occurs.}
\end{figure}

\begin{figure}
\begin{center}
\includegraphics[angle=0, width=0.40\textwidth]{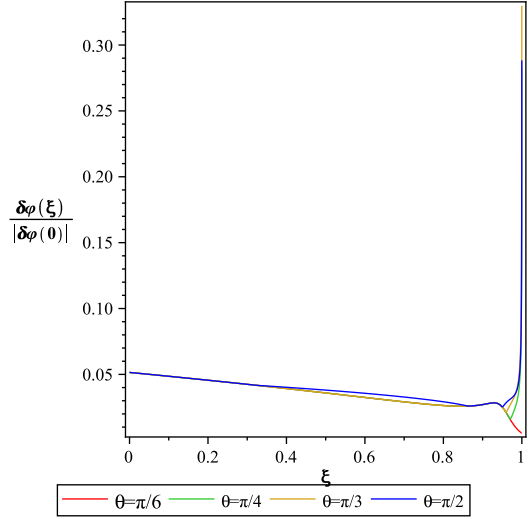}
\end{center}
\caption{\label{Fig3a} Evolution of Nariai horizon perturbation for $n=1$ and different choices of $\theta$ in the model $F(R)=R+\gamma R^{10}$. The horizon grows up and finally diverges. Anti-evaporation process finally occurs, but it is possible to observe some transient evaporation effects for small times.}
\end{figure}

Let us explicitly see how Nariai solution for $n=1$ mode perturbation evolves depending on $\mu$. 
In Figs.~\ref{Fig2a}, \ref{Fig3a}, the cases of real values of $\mu$ are shown: we depicted the evolution of the horizon in the model $F(R)=R+\gamma R^m$ (\ref{inflmodel}), which posses unstable SdS solution for $\gamma>0$ and $m>2$, by choosing $m=5$ (namely, $\alpha=5/4$ and $\mu=2/\sqrt{5}$) and $m=10$ (namely, $\alpha=5/2$ and $\mu=4/\sqrt{15}$), respectively.
In order to evaluate the perturbation (\ref{pert}), we wrote the parameters $a\,,b$ as in (\ref{ab})
and we made different choices of $\theta$, namely $\theta=\pi/6\,,\pi/4\,,\pi/3\,,\pi/2$.
The perturbation has to be normalized on $|\delta\varphi(0)|$. One should remember that the relation between $\delta\varphi$ and Nariai horizon is given by (\ref{relation}). We can see that in the both cases the Nariai solution is unstable and the horizon grows up and diverges, producing a final anti-evaporation of the black hole. Note that before the final evaporation,
some transient effects of evaporation are present.
As we noted above, if $\epsilon$ is negative, we expect that these black holes finally evaporate.

\begin{figure}
\begin{center}
\includegraphics[angle=0, width=0.40\textwidth]{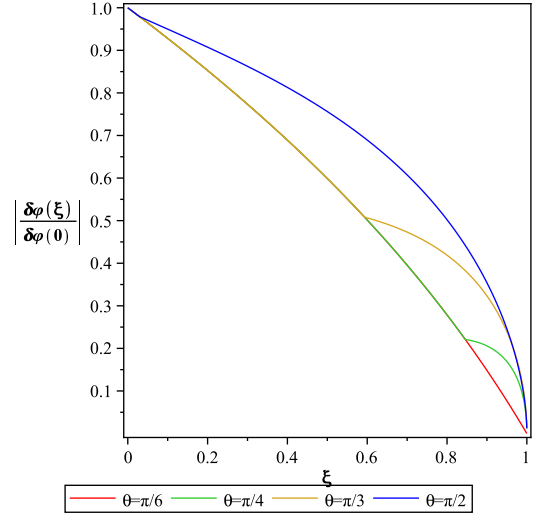}
\end{center}
\caption{\label{Fig4a} Evolution of Nariai horizon perturbation for $n=1$ and different choices of $\theta$ in the model $F(R)=R+\gamma R^8$. Despite the fact that the model under consideration posseses an unstable de Sitter solution, the corresponding Nariai solution results to be stable, namely the intensity of perturbations descreases and finally tends to zero.}
\end{figure}

In Fig.~\ref{Fig4a} we show the special case of $F(R)=R+\gamma R^8$ ($m=8$), $\alpha=2$, which corresponds to $\mu=1$, namely it is an integer number. In this case, the solution is stable and the intensity of perturbation (we depicted $|\delta\varphi(\tau)/\delta\varphi(0)|$) decreases and finally disappears according with (\ref{according}).

\begin{figure}
\begin{center}
\includegraphics[angle=0, width=0.40\textwidth]{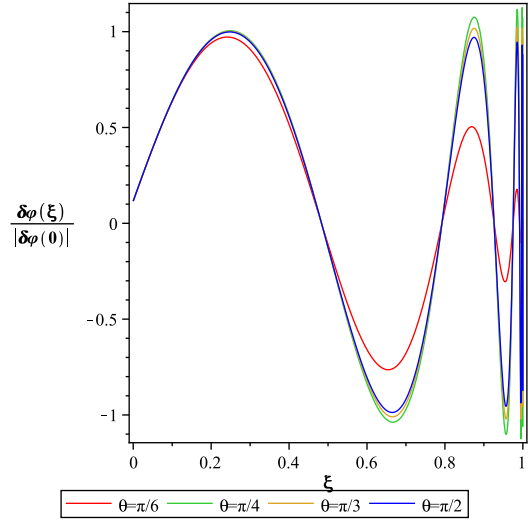}
\end{center}
\caption{\label{Fig2} Evolution of Nariai horizon perturbation for $n=1$ and different choices of $\theta$ in the Hu-Sawicki model (\ref{model2}). Here, the Nariai solution results to be stable.}
\end{figure}

\begin{figure}
\begin{center}
\includegraphics[angle=0, width=0.40\textwidth]{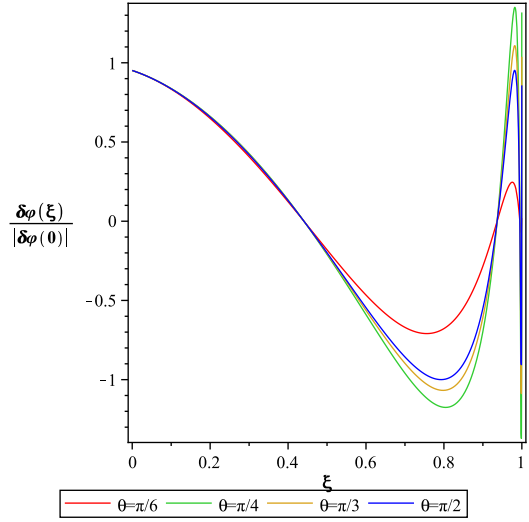}
\end{center}
\caption{\label{Fig3}Evolution of Nariai horizon perturbation for $n=1$ and different choices of $\theta$ in exponential gravity (\ref{model}). Here, the Nariai solution results to be stable.}
\end{figure}

In Figs.~\ref{Fig2}, \ref{Fig3} the cases of imaginary values of $\mu$ are shown: we depicted the evolution of the horizon in the Hu-Sawicki model ($\alpha=0.02$) and in exponential gravity ($\alpha=0.09$), respectively. We wrote again the parameters $a\,,b$ as in (\ref{ab})
and we put $\theta=\pi/6\,,\pi/4\,,\pi/3\,,\pi/2$.
The perturbation has to be normalized on $|\delta\varphi(0)|$ again and the real part has been taken. One can see that in the both cases the Nariai solution is an attractor, and the evaporation/anti-evaporation phases are only transient effects.

\subsubsection{Higher modes perturbations}

\begin{figure}
\begin{center}
\includegraphics[angle=0, width=0.40\textwidth]{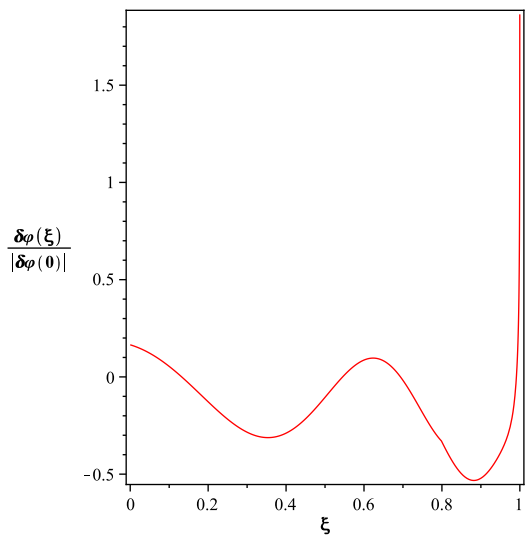}
\end{center}
\caption{\label{Fig33}Evolution of Nariai horizon perturbation for n=5 and $\theta=\pi/4$ in the model $F(R)=R+\gamma R^5$. After some transient phases of evaporation and anti-evaporation, the horizon grows up and diverges, giving a final anti-evaporation of the black hole.}
\end{figure}

\begin{figure}
\includegraphics[angle=0, width=0.40\textwidth]{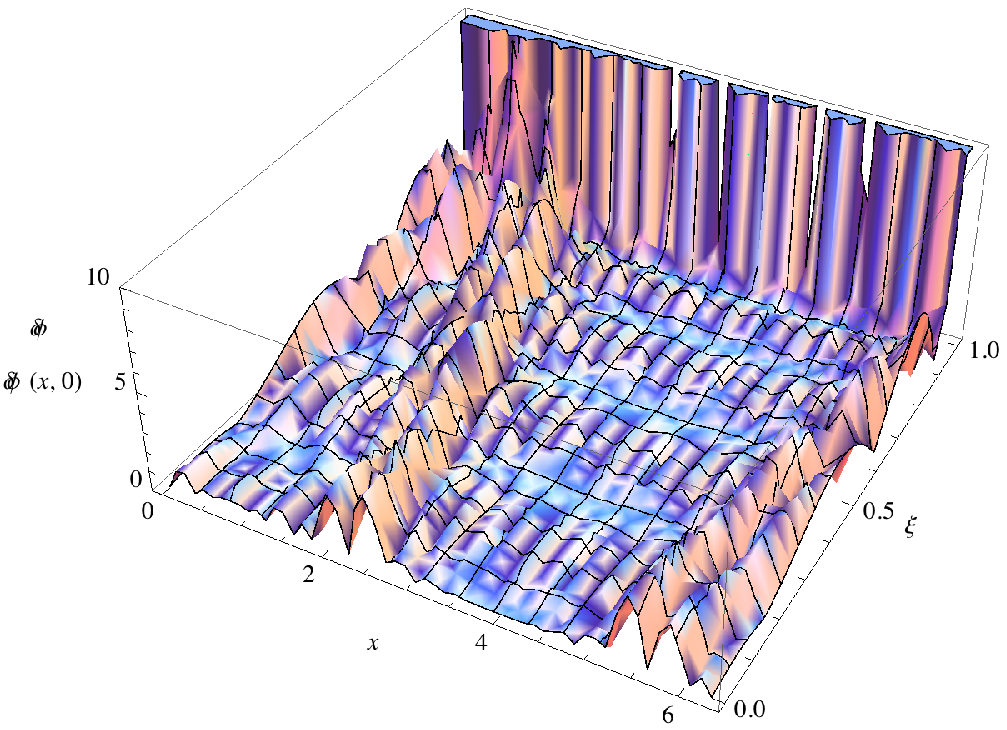}
\quad
\includegraphics[angle=0, width=0.40\textwidth]{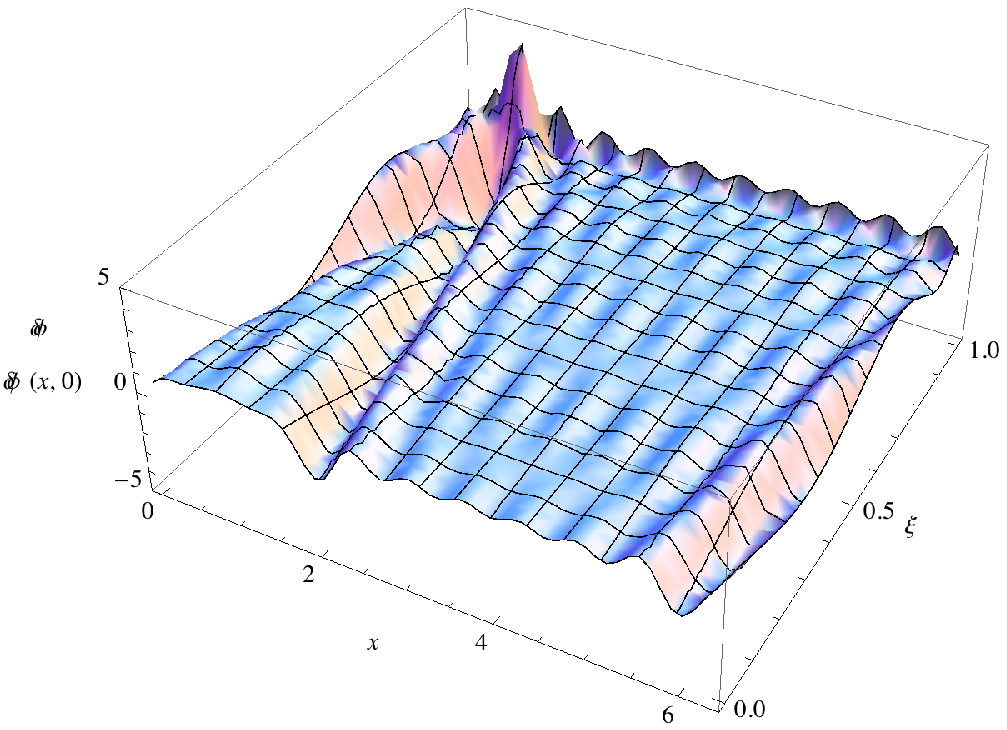}
\caption{\label{Fig3333}Evolution of Nariai perturbation as function of $x$ and $\xi (=\sin\tau)$ for superposition of different modes and $\theta=\pi/4$ in two different models, namely  the model $F(R)=R+\gamma R^{10}$ in the left panel and in the exponential model (\ref{model}) in the right panel. In the first case, perturbations diverge, in the second they oscillate around the Nariai solution.}
\end{figure}

The analysis for a single mode $n>1$ in (\ref{finalpert}) is not so different from the one carried out in the previous Subsection. For simplicity, we can put
\begin{equation}
a_n=\cos n\theta\,,\quad b_n=\sin n\theta\,,\label{53}
\end{equation}
where $\theta$ is a fixed angle, and, as a consequence, perturbation reads
\begin{equation}
\delta\varphi=\epsilon P^{\nu}_{\mu}(\xi)\cos\left[n(\theta-x)\right]\,,\quad\nu=-\frac{1}{2}\pm\sqrt{n^2+\frac{1}{4}}\,.\label{single}
\end{equation}
Again, we will choose the plus sign for $\nu$.
On the horizon, 
\begin{equation}
\cos \left[n(\theta-x)\right]=\pm\sqrt{\frac{1-\xi^2}{n^2}\left(\frac{d P_\nu^\mu(\xi)}{d\xi}\frac{1}{P^\mu_\nu(\xi)}\right)^2+1}\,.
\end{equation}
Therefore, the perturbation can be written as (with the plus sign for the cosinus) 
\begin{eqnarray}
\delta\varphi(\xi)=\epsilon\,P^{\mu}_{\nu}(\xi)\left\{
\sqrt{\frac{1-\xi^2}{n^2}\left(\frac{d P_\nu^\mu(\xi)}{d\xi}\frac{1}{P^\mu_\nu(\xi)}\right)^2+1}
\right\}\,.
\end{eqnarray} 
Also in this case, the perturbation is stable when $0<\alpha<1/2$ (i.e. $\mu$ is imaginary) and diverges when $\alpha<0$ or $1/2<\alpha$ (i.e. $\mu$ is real), since when $\xi$ is close to $1^-$, namely in the future, the Legendre polynomial for generic $\nu$ expands as
\begin{eqnarray}
P_{\nu}^{\mu}(\xi)=(-1)^{\frac{\mu}{2}}(1-\xi)^{-\frac{\mu}{2}}\Big(\frac{2^{\frac{\mu}{2}}}{\Gamma(1-\mu)}-2^{\frac{\mu}{2}-2}(1-\xi)\Big)
\left(\frac{\mu}{\Gamma(1-\mu)}+\frac{2\nu(1+\nu)}{\Gamma(2-\mu)}\right)(1-\xi)+\mathcal{O}(1-\xi)^2\,,&&\nonumber\\
(1-\xi)\ll 1\,,
\end{eqnarray}
in analogy with Eq.~(\ref{mureal}) and Eq.~(\ref{50}).
As an example, in Fig. (\ref{Fig33}) we depict the case of $\alpha=5/4$ (it corresponds to the model (\ref{inflmodel}) with $m=5$) with $\theta=\pi/4$.
It is interesting to note that for higher mode perturbation (in this case we plotted $n=5$) the mode starts to oscillate until its intensity grows up enough to leave the horizon and diverge. It means that before the final anti-evaporation, there is a proliferation of evaporation/anti-evaporation phases.
This process was firstly observed in Ref.~\cite{Nie} by considering 2d quantum instabilities in Nariai black holes. 
\\ \\
Let us consider the more general case (\ref{finalpert}). We write the coefficients ${a_n,b_n}$ as in (\ref{53}), such that for the horizon perturbation we get
 \begin{eqnarray}
 \delta\varphi(x,t)=\epsilon\sum_{n=1}^{N}P_{\nu}^{\mu}(\xi)\cos\left[n(\theta-x)\right]\,,\label{58}
 \end{eqnarray} 
where $N\gg 1$. One also recovers the single case mode (\ref{single}). In this case, we cannot solve Eq.~(\ref{horizon}) for the location of Nariai horizon. However, the stability of the solution still depends on the behaviour of Legendre polynomials, and we can predict a final evaportation or anti-evaporation of the black hole only when $\alpha<0$ or $1/2<\alpha$, namely for $F(R)$-gravity which admits unstable dS-solution. In Fig. (\ref{Fig3333}) we depict the evolution of the absolute value of perturbation (\ref{58}) as function of $x$ and $\tau$ for $\alpha=5/2$ (it corresponds to the model $F(R)=R+\gamma R^m$ (\ref{inflmodel}) with $m=10$) in the left panel and the perturbation for $\alpha=0.09$ (it corresponds to the exponential model (\ref{model})) in the right panel. We set $N=10$ and $\theta=\pi/4$. Furthermore, we have normalized $\delta\varphi(x,\xi)$ to $|\delta\varphi(x,0)|$. In the first case, we can see that perturbations grow up and diverge in time (almost for every value of $x$), and in the second case the perturbations oscillate and never leave the horizon.

\section{Nariai solution in the static patch of $F(R)$-gravity}

\begin{figure}
\includegraphics[angle=0, width=0.40\textwidth]{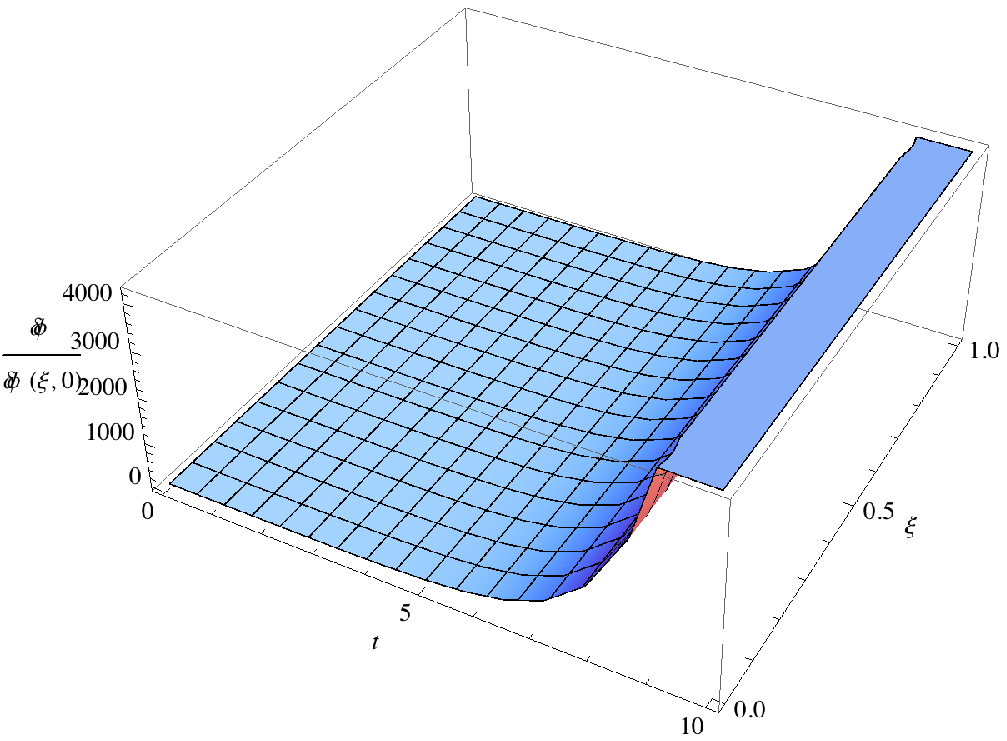}
\quad
\includegraphics[angle=0, width=0.40\textwidth]{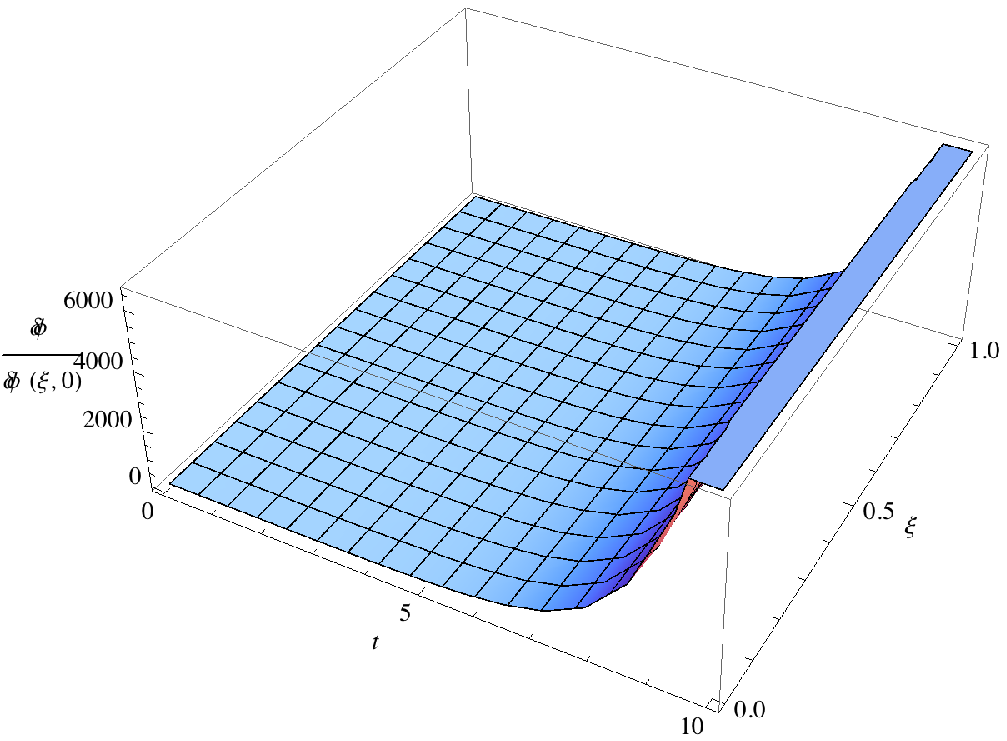}
\caption{\label{Fig0}Evolution of Nariai perturbation as function of $\xi(=\tanh x)$ and $t/T$ in the static patch for $\omega=i T$ and $\theta=\pi/4$ in
Hu-Sawicki model (left side) and exponential gravity (right side). In the both cases, perturbations diverge.}
\end{figure}

In this Section, the analysis of instabilities in the static patch of $F(R)$-Nariai black holes is done. 
Note that static patch description is usually believed to be less complete one.
The static patch of Nariai solution is given by the metric (\ref{Nr1}), from which one can easily derive
\begin{equation}
ds^2=\frac{1}{\Lambda\,\cosh^2 x}\left(-d t^2+d x^2\right)+\frac{1}{\Lambda}d\Omega^2\,,
\end{equation}
where $x=\cosh^{-1}\left[1/\sin\chi\right]$ such that $-\infty<x<+\infty$ and $t(\equiv\psi)$ is the time coordinate (in the following, the dot will denote the derivative with respect to $t$). For this form of the metric, we can still use the Ansatz (\ref{Nr3}) and the perturbations on Nariai metric $\delta\rho(t,x)\,,\delta\varphi(t,x)$ can be written as
\be
\label{newNr6}
\rho = - \ln \left[ \sqrt{\Lambda} \cosh x \right] + \delta\rho\, ,\quad
\varphi = \ln \sqrt{\Lambda} + \delta \varphi\,.
\ee
For the Ricci scalar perturbation we get
\be
\delta R = 4 \Lambda \left( - \delta\rho + \delta\varphi \right)
+ \Lambda \cosh^2 x \left( 2 \delta \ddot\rho - 2 \delta \rho''
 - 4 \delta \ddot\varphi + 4\delta \varphi'' \right) \, .
\ee
The perturbed equations of motion have been derived in Ref. \cite{Od} by starting from (\ref{Nr5}) and may be obtained by replacing $\cos \tau$ with $\cosh x$ and $\tan\tau$ with $-\tanh x$ in the system (\ref{Nr8}). Also in the static patch we finally deal with two equations. The first one corresponds to Eq.~(\ref{deltaR}) with $C_x(x)/\cos\tau\rightarrow C_x(x)$ and $C_\tau(\tau)\rightarrow C_t(t)/\cosh x$ and we can put $C_x(x)=C_t(t)=0$ again. As a consequence, the second equation reads~\cite{Od}
\begin{equation}
\frac{1}{\alpha\cosh^2 x}\left[2(2\alpha-1)\delta\phi\right]-3\delta\ddot\varphi+3\delta\varphi''=0\,,
\label{principe2}
\end{equation}
where $\alpha$ is still given by (\ref{alpha}). The last equation can be used to study the evolution of the horizon as in (\ref{relation}). 
By introducing $\xi=\tanh x$ such that $-1<\xi<1$, 
we can check for the solutions of the above expression in the following form
\begin{equation}
\delta\phi=\epsilon\Big[a_{\omega}P_{\nu}^{i\omega}(\xi)\cos\omega t+b_{\omega}P_{\nu}^{-i\omega}(\xi)\sin\omega t\Big]\,,\quad1\gg\epsilon>0\,,\label{ooo}
\end{equation}
where $\epsilon$ is positive and small, and $\omega$ is a frequency number which may assume real or complex values. Furthermore, $\nu$ results to be
\begin{equation}
\quad\nu=\frac{-3\pm\sqrt{3\left(19-\frac{8}{\alpha}\right)}}{6}\,,
\end{equation}
and the coefficients $a_\omega$ and $b_\omega$ can be choosen as
\begin{equation}
a_\omega=\cos\theta\,,\quad b_\omega=\sin\theta\,,
\end{equation}
where $\theta$ is an angular parameter.
The above expression can be derived in an analogous way of the ones for the cosmological patch.
A more general solution of (\ref{principe2}) is given by a superposition of (\ref{ooo}) with different values of $\omega$, but here we will restrict our discussion to the case of single $\omega$.
From Eq. (\ref{horizon}) we have the position of the two-sphere,
\begin{equation}
\tan\omega t=\pm\frac{-(1-\xi^2)a_\omega \frac{ d P^{i\omega}_\nu(\xi)}{d\xi}+\omega b_\omega P_\nu^{-i\omega}(\xi)}{(1-\xi^2)b_\omega\frac{ d P^{-i\omega}_\nu(\xi)}{d\xi}+\omega a_\omega P^{i\omega}_\nu(\xi)}\,,
\end{equation}
but in general it is not possible to analytically solve it with respect to $\xi$. 
However, one can easily see that we have no restriction on $\omega$ and when the frequency becomes imaginary, the solution diverges and becomes unstable independently of the model. As the examples, in Fig.~(\ref{Fig0})
we plot the 
the evolution of the real part of perturbation (\ref{ooo}) as function of $\xi$ and $t$ for
Hu-Sawicki model ($\alpha=0.02$) in the left panel and the exponential gravity ($\alpha=0.09$)
in the right panel for the $\omega=iT$, where $T$ is a frequency parameter (the time is normalized on $T$). We set $\theta=\pi/4$
and we depict $|\delta\varphi(\xi,t)/\delta\varphi(\xi,0)|$. 
In both of the cases, the perturbation grows up in intensity and diverge almost for every value of $\xi$ (and therefore of the comoving coordinate $x$). Remind that for this class of models that provides stable Schwarzschild-dS solution, Nariai perturbations are stable in the cosmological patch, but if we consider the static one new feature may arise.
In fact, we observe that in the cosmological patch the $S_1$ sphere expands exponentially according with the metric, but in the
static patch the time dependence disappears. As a consequence, it may
happen that some perturbations in the static patch diverge, but in the
cosmological patch they grow up with the $S_1$ sphere and appear to be stable.
Since the cosmological patch is considered to be more general description of Nariai black hole, 
we may argue that perturbation analysis which we did in this patch is more completed.

\section{Nariai black holes in Gauss-Bonnet modified gravity}

In this Section, in the attempt to study the Nariai black holes in a more general class of modified theories, we will investigate the case of $f(G)$-gravity (for black hole solutions in $f(G)$-gravity, see Refs.~\cite{SSSFG}).
We start from the following action~\cite{fG},
\begin{eqnarray}
I=\int_{\mathcal M} d^4x\sqrt{-g}\Big[\frac{R+f(G)}{2\kappa^2}\Big]\,,
\end{eqnarray}
such that the modification to gravity is given by the function $f(G)$
of the Gauss-Bonnet four-dimensional topological invariant
\begin{equation}
G=R^{2}-4R_{\mu\nu}R^{\mu\nu}+R_{\mu\nu\xi\sigma}R^{\mu\nu\xi\sigma}\,.\label{GaussBonnet}
\end{equation}
The Gauss-Bonnet invariant is a combination of the Riemann tensor $R_{\mu\nu\xi\sigma}$, the Ricci tensor $R_{\mu\nu}=R^{\rho}_{\mu\rho\nu}$ and its trace $R=g^{\alpha\beta}R_{\alpha\beta}$.
As in the case of $F(R)$-gravity, we ignore the matter contribution. The field equations are
\begin{eqnarray}
&&R_{\mu\nu}-\frac{1}{2}Rg_{\mu\nu}=\frac{1}{2}g_{\mu\nu}f-2FRR_{\mu\nu}+4FR_{\mu\rho}R_{\nu}^{\rho}
\nonumber\\ \nonumber\\
&&-2FR_{\mu}^{\rho\sigma\tau}R_{\nu\rho\sigma\tau}-4FR_{\mu\nu}^{\rho\sigma}R_{\rho\sigma}+2R\nabla_{\mu}\nabla_{\nu}f'-2Rg_{\mu\nu}\nabla^2 F\nonumber\\ \nonumber\\
&&-4R_{\nu}^{\rho}\nabla_{\rho}\nabla_{\mu}F-4R_{\mu}^{\rho}\nabla_{\rho}\nabla_{\nu}F+4R_{\mu\nu}\nabla^2F+4g_{\mu\nu}R^{\rho\sigma}\nabla_{\rho}\nabla_{\sigma}F-4R^{\rho\sigma}_{\mu\nu}\nabla_{\rho}\nabla_{\sigma}F.
\label{GB-EOM}
\end{eqnarray}
Here, we use the following notation:
\begin{equation}
F=\frac{\partial f(G)}{\partial G}\,.
\end{equation}
Let us assume that our Gauss-Bonnet model admits the Nariai solution (\ref{Nariai}) for $R_0=4\Lambda$ and
$G_0=2\Lambda^2$. It is interesting to note that the Gauss-Bonnet invariant on Schwarschild-dS solution depends on the radial coordinate ($G=(2/3)[31\Lambda^2+18 M^2/r^6]$). However, it is constant on Nariai solution (in fact, the radial coordinate corresponds to the radius of Nariai horizon). 
Moreover, in order to obtain the Nariai solution, it is enough that the model admits the de Sitter space-time
for $R=R_0\,,G=G_0$.
In this analysis, we will work with the cosmological patch of the solution. From the trace of field equations we obtain the following condition
\begin{eqnarray}
R_0=-2f+2FG_0\,,\label{traceGB0}
\end{eqnarray}
that the model must satisfy in order to admit the Nariai solution.
We assume the metric Ansatz (\ref{Nr3}).
Now, we consider the perturbation on the Nariai space-time in the form of
(\ref{Nr6}). The perturbation of the Gauss-Bonnet invariant results to be at the first order (see the Appendix for the complete form of the Gauss-Bonnet invariant)
\begin{eqnarray}
\delta G=G_0\left[\delta\ddot{\rho}-\frac{5}{2}\cos^2\tau\delta\rho''+2\cos^2\tau\delta\varphi+2(\cos^2\tau-2)\delta\rho\right]\,.\label{deltaG}
\end{eqnarray}
Since the trace of the field equations (\ref{GB-EOM}) reads
\begin{eqnarray}
-R=2f-2FG-2R\Box F+4R_{\mu\nu}\nabla^{\mu}\nabla^{\mu}F\,,\label{traceGB}
\end{eqnarray}
we obtain at the first order
\begin{eqnarray}
\delta R=2F'(G_0)\Big(G_0\delta G+R_0\Box \delta G-2R^{0}_{\mu\nu}\nabla^{\mu}\nabla^{\nu}\delta G\Big)\,.
\end{eqnarray}
Here, $R^{0}_{\mu\nu}$ denotes the Ricci tensor of Nariai metric and $\delta R$ is given by (\ref{varR}). This expression leads to
\begin{eqnarray}
\frac{\delta R}{4\Lambda^2 F'(G_0)}=\delta G+\cos^2\tau(\delta\ddot{G}-\delta G'')\label{deltaGR}.
\label{deltaG}
\end{eqnarray}
The perturbed field equations (see the Appendix for their complete form) read
\bea
&&4\Lambda(F(G_0)+\Lambda\cos^2\tau)\delta\ddot{\rho}+\Big(2
+8\Lambda(F(G_0)\sec^2\tau-\Lambda\cos^2\tau)\Big)\delta\ddot{\varphi}
+\Big(1-4\Lambda(F(G_0)+\Lambda\cos^2\tau)\Big)\delta\rho''
\nonumber\\\nonumber\\&&
-2\tan\tau\Big(1+4\Lambda F(G_0)-4\Lambda(F(G_0)\sec^2\tau+\Lambda\cos^2\tau)\Big)\delta\dot{\varphi}+ \sec^2\tau\Big(\frac{1}{2\Lambda}-2F(G_0)\Big)\delta R
\nonumber\\\nonumber\\&&
+\delta\rho\Big(\sec^2\tau(4+\frac{f(G_0)}{\Lambda}-12\Lambda F(G_0))-8\Lambda^2\Big)
+ \sec^2\tau\Big(\frac{F(G_0)}{2\Lambda}-2\Lambda F'(G_0)\Big)\delta G+8\Lambda F'(G_0)\delta \dot{G'}
\nonumber\\\nonumber\\&&
+4\Lambda F'(G_0)\delta{\ddot{G}}
-8\Lambda\tan\tau F'(G_0)\delta\dot{G}+4\Lambda F'(G_0)\delta G''-8\Lambda\tan\tau F'(G_0)\delta G'+8\Lambda^2(\delta\varphi-\delta\rho)=0\,,\nonumber\\\label{pert-tt}
\eea
\bea
&&(-1+4\Lambda F(G_0))\delta\rho''+(1+4\Lambda F(G_0))\delta\ddot{\rho}+2\delta\varphi''-2\tan\tau (1-4\Lambda F(G_0))\delta\dot{\varphi}
\nonumber\\\nonumber\\&&
+\sec^2\tau\Big(-4-\frac{f(G_0)}{\Lambda}+4\Lambda F(G_0)\Big)\delta\rho+
 \sec^2\tau\Big(2F(G_0)-\frac{1}{2\Lambda}\Big)\delta R
\nonumber\\\nonumber\\&&
+ \sec^2\tau\Big(-\frac{F(G_0)}{2\Lambda}+6\Lambda F'(G_0)\Big)\delta G+4\Lambda F'(G_0)( \delta G''-\delta\ddot{G})=0\,,\label{pert-xx}
\eea
\bea
&&\Big(1+4\Lambda F(G_0) (3-2\cos^2\tau)\Big)\delta\dot{\varphi}'-\tan\tau\Big(1+8\Lambda F(G_0)\sin^2\tau\Big)\delta\varphi'-6\Lambda F'(G_0)\Big(\delta\dot{G'}-\tan\tau\delta G'\Big)=0\label{pert-xt}\,,\nonumber\\
\eea
\bea
&&-\cos\tau\Big(\cos\tau+4\Lambda F(G_0)(1-\cos^3\tau)\Big)\delta\ddot{\varphi}+\cos\tau\Big(\cos\tau+4\Lambda F(G_0)\Big)\delta\varphi''+\Big(2F(G_0)-\frac{1}{2\Lambda}
\nonumber\\\nonumber\\&&
-\frac{4 F(G_0)}{\Lambda}\Big)\delta R(\frac{4}{\Lambda}+\frac{f(G_0)}{\Lambda}+\frac{4\Lambda F(G_0)}{\cos\tau})\delta\varphi+\Big(8\Lambda F'(G_0)-\frac{4F'(G_0)}{\Lambda}-2\Lambda F'(G_0)\sec^2\tau-\frac{F(G_0)}{2\Lambda}\Big)\delta G
\nonumber\\\nonumber\\&&
+4\Lambda F(G_0)\sec\tau (1-\sec\tau)\delta\rho-4\Lambda F(G_0)\cos^4\tau\tan\tau \delta\dot{\varphi}=0\,.
\eea
By integrating the third equation (\ref{pert-xt}), we obtain
\bea
\delta\varphi'=\mu(\tau)\Big[C_x(x)+6\Lambda F'(G_0)\int{\frac{\partial(\cos\tau \delta G') }{\partial\tau}\frac{d\tau}{\mu(\tau)\Big((1+9\Lambda F(G_0))\cos\tau-\Lambda F(G_0)\cos(3\tau)\Big)}}\Big]\,,\nonumber\\\label{firstintegral}
\eea
where
\bea
\mu(\tau)=\Big[\cos\tau\Big]^{-\frac{1+8\Lambda F(G_0)}{1+12\Lambda F(G_0)}}\Big[1+2\Lambda F(G_0)(5-\cos(2\tau))\Big]^{-\frac{1+16\Lambda F(G_0)}{2(1+12\Lambda F(G_0))}}\,.
\eea
By integrating (\ref{firstintegral}) again one has
\bea
\delta\varphi=\mu(\tau)\Big[C_x(x)+6\Lambda F'(G_0)\int{\frac{\partial(\cos\tau \delta G) }{\partial\tau}\frac{d\tau}{\mu(\tau)\Big((1+9\Lambda F(G_0))\cos\tau-\Lambda F(G_0)\cos(3\tau)\Big)}}\Big]+C_\tau(\tau)\,.\nonumber\\\label{secondintegral}
\eea
For simplicity we can put $C_x(x)=C_\tau(\tau)=0$ and finally we obtain the following integro-differential equation for horizon perturbation $\delta\varphi$ versus Gauss-Bonnet perturbation $\delta G$:
\bea
\delta\varphi=6\Lambda F'(G_0)\mu(\tau)\int{\frac{\partial(\cos\tau \delta G) }{\partial\tau}\frac{d\tau}{\mu(\tau)\Big((1+9\Lambda F(G_0))\cos\tau-\Lambda F(G_0)\cos(3\tau)\Big)}}\,.\label{deltaphi-sol}
\eea
As we did for the case of $F(R)$-gravity, we decompose horizon perturbation, and therefore the Gauss-Bonnet perturbation, in Fourier modes as\\
\phantom{line} 
\bea
\delta G=\sum_{n=1}^{\infty}(\alpha^{c}_n(\tau)\cos(nx)+\alpha^{s}_{n}(\tau)\sin(nx))\,,\quad
\delta\varphi=\sum_{n=1}^{\infty}(\beta^{c}_n(\tau)\cos(nx)+\beta^{s}_{n}(\tau)\sin(nx))\label{phi-fourier}.
\eea
\phantom{line}\\
By plugging this expressions in (\ref{deltaphi-sol}), we derive the following equations for Fourier components $X_n(\tau)=\{\alpha^{c,s}_n(\tau),\beta^{c,s}_n(\tau)\}$:\\
\phantom{line}
\bea
\beta^{c}_n(\tau)=6\Lambda F'(G_0)\mu(\tau)\int{\frac{\partial(\cos\tau\, \alpha^{c}_n(\tau)) }{\partial\tau}\frac{d\tau}{\mu(\tau)\Big((1+9\Lambda F(G_0))\cos\tau-\Lambda F(G_0)\cos(3\tau)\Big)}}\,,\nonumber\\
\beta^{s}_n(\tau)=6\Lambda F'(G_0)\mu(\tau)\int{\frac{\partial(\cos\tau\, \alpha^{s}_n(\tau)) }{\partial\tau}\frac{d\tau}{\mu(\tau)\Big((1+9\Lambda F(G_0))\cos\tau-\Lambda F(G_0)\cos(3\tau)\Big)}}\label{betas}\,.
\eea
\phantom{line}\\
For the first mode $n=1$, from the second expression in (\ref{phi-fourier}), one derives the location of the horizon (\ref{horizon}),
\bea
\tan x=\frac{\beta_1^s-\beta_1^{c'}}{\beta_1^{s'}+\beta_1^{c}}\,,
\eea
such that the horizon perturbation reads
\bea
\delta\varphi(\tau)=\frac{\beta_1^c(\tau)(\beta_1^{s'}(\tau)+\beta_1^{c}(\tau))+\beta_1^s(\tau)(\beta_1^s(\tau)-\beta_1^{c'}(\tau))}{\sqrt{(\beta_1^s(\tau)-\beta_1^{c'}(\tau))^2+(\beta_1^{s'}(\tau)+\beta_1^{c}(\tau))^2}}\,\label{deltaphi-GB}.
\eea
We remind that $\beta_1^{c,s}(\tau)$ encode the Gauss-Bonnt model that realizes the Nariai solution as in (\ref{betas}).\par
 Working with (\ref{betas}) is very complicated, since we must solve a system of four coupled differential equations. Therefore, in order to obtain a full description of the horizon perturbation, we must also know the coefficients $\{\alpha^{c}_n(\tau),\alpha^{s}_n(\tau)\}$. One possibility is given by the numerical analysis, but also in this case we need to specify the boundary conditions of fields on a suitable Cauchy surface. Here, instead of the very complicated numerical analysis, we will make use of a simple Ansatz for $\delta G$, inspired from the form of Gauss-Bonnet invariant on the Nariai solution. 
The relation between the Ricci scalar and the Gauss-Bonnet invariant on Nariai space-time is given by
\begin{equation}
G_0=\frac{R_0^2}{8}\,.
\end{equation}
Inspired from this relation, we may assume the following Ansatz for Ricci and Gauss-Bonnet perturbations 
\begin{equation}
\frac{\delta R}{4\Lambda^2 F'(G_0)}=(1+\gamma)\delta G\,,
\end{equation}
where $\gamma=\frac{1-m^2}{4},m\in[0,1]$ , $0\leq \gamma<\frac{1}{4} $ is a dimensionless parameter. With this simple Ansatz we can integrate Eq.~(\ref{deltaG}) which leads to
\begin{eqnarray}
\{\alpha^{c}_n(\tau),\alpha^{s}_n(\tau)\}=c_{n}^{\pm}(\cos\tau)^{\frac{1\pm m}{2}} F\left[\frac{1\pm m-2n}{4},\frac{1\pm m+2n}{4},1\pm\frac{m}{2};\cos^2\tau\right]\label{umn}\,,
\end{eqnarray}
where $F(a,b,c;z)$ represents, as usually, the hypergeometric series and $c^\pm$ are constants. Thus, we can set the Fourier components and find the horizon perturbation using the integral representations in (\ref{betas}).

\begin{figure}
\begin{center}
\includegraphics[angle=0, width=0.4\textwidth]{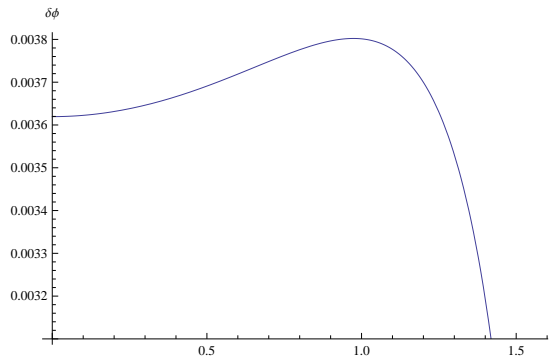}
\end{center}
\caption{\label{Fig22222} Evolution of horizon perturbation for the $f(G)$-model (\ref{model}) with $\lambda=10^{12}\,,m=0.25$ and $G_s=H_0^4$, $H_0$ being the Hubble parameter in the accelerated universe. We note that perturbation grows up until a tourning point, after that decreases and tends to zero in a finite time.}
\end{figure}

Now, as a specific example,
we need a ``viable" toy-model of $f(G)$-gravity. 
We use the one proposed in Ref.
~\cite{fg-obs}, which is consistent with the observational data in accelerated universe, namely 
\bea
f(G)=\lambda\sqrt{G_s}\Big[-\alpha+g(x)\Big]\,,\quad x=\frac{G}{G_s},\quad g(x)=x\arctan x-\frac{1}{2}\log(1+x^2)\label{model}\,.
\eea
Here ${\alpha,\lambda}$ are real positive constant parameters and $G_s\sim H_0^4$, where $H_0$ denotes the Hubble parameter in the de Sitter universe. 
From different observational constraints, like solar system tests, Cassini experiment and so on, one finds that $\lambda\sim10^5-10^{15}$. 
From (\ref{model}), one has
\bea
F(G_0)=\frac{\lambda}{\sqrt{G_s}}\arctan\Big[\frac{2\Lambda^2}{G_s}\Big]\,,\quad F'(G_0)=\frac{\lambda}{\sqrt{G_s}}\Big[1+\left(\frac{2\Lambda^2}{G_s}\right)^2\Big]^{-1}\,.
\eea
and, by taking into account that $H_0=\sqrt{\frac{\Lambda}{3}}$, $\Lambda$ being the cosmological constant, 
\bea
F(G_0)=\frac{4.54589\lambda}{\Lambda}\,,\quad F'(G_0)=\frac{0.00923\lambda}{\Lambda}\,,
\eea
when $G_s=H_0^4$.
This model admits the Nariai solution for $R=4\Lambda\,, G=2\Lambda^2$.
Now we can compute
(\ref{betas}) with (\ref{umn}). We put $c^\pm=(2\pi)^{-1/2}$. We will limit to the analysis of the first mode, namely $n=1$, 
which is the longest wavelength. Moreover, for the sake of simplicity, 
we will consider only the odd horizon and Gauss-Bonnet perturbations. It means that we have 
set $\alpha^{s}_{n}(\tau)=0\,,\beta^{s}_{n}(\tau)=0$ in (\ref{phi-fourier}). 
So, by integrating (\ref{betas}) and plugging the result in (\ref{deltaphi-GB}), we obtain the horizon perturbation. The related graph for $\lambda=10^{12}$ and $m=0.25$ is shown in Fig.~\ref{Fig22222}.    
In this case the
final evolution assumes a stable configuration and the perturbation tends to zero. Before the final stable point at $\tau=1.5708$, we have an anti-evaporation phase ($0<\tau<\tau^*$) following
by an evaporation phase. The value 
$\tau^{*}\simeq 0.922168$ corresponds to a turning point between the two phases.
We do not have
a pure evaportation/anti-evaporation as the final evolution. Near to the turning point, the system becomes unstable and, due to an infintesimal deviation
from the maxima, falls in an evaporation phase which brings the Nariai black hole to its initial size ($\delta\varphi=0$). As a consequence, in this specific example of $f(G)$-gravity, the Nariai black hole results to be stable.
Of course, other models of $f(G)$-gravity maybe analyzed in the same way. However, it turns out that the corresponding analysis is much more complicated than in $f(R)$-gravity.

\section{Discussion}

In summary, we investigated the evolution of Nariai black holes in $F(R)$-gravity. The metric of the extremal limit of Schwarzschild-de Sitter black holes 
is considered in cosmological and static patches. The perturbation equations 
are presented in both patches. It is indicated that horizon perturbations 
depend on the cosmological time and on the comoving coordinate $x$. In the 
cosmological patch the horizon perturbations are decomposed into Fourier modes 
whose coefficients are expressed as the Legendre polynomials. The study of 
$n=1$ mode, generic modes as well as superposition of different modes is made. 
It turns out that when SdS solution in $F(R)$-gravity is stable, also its 
extremal limit results to be stable. However, when SdS solution is not stable, 
the extremal limit maybe stable for some specific cases. Furthermore, when the 
solution is unstable, not only black hole evaporation but also its 
anti-evaporation may occur. These considerations are applied to several models 
of $F(R)$-gravity which describe current dark energy epoch or early-time 
inflation.  The corresponding analysis is done analytically and numerically. It 
turns out that Nariai black holes for some of the models under discussion may 
enter to an unstable phase or even anti-evaporate.
Hence, the presence of Nariai black holes at the current epoch may favour the 
alternative gravity which supports such anti-evaporation. Even more, as current 
dark energy gravity may also support such anti-evaporation, one can speculate 
about the possibility to get huge black holes in the future universe just before 
the Rip occurence.

In the static patch of Nariai solution some new feature may appear. Namely, 
even if the model admits stable SdS solution, the Nariai black hole results to be 
unstable. Hence, the (anti)-evaporation regions maybe different in static patch 
if compared with cosmological patch. This is not surprising. Indeed, it is known 
that cosmological patch description is considered to be more complete than 
static patch description. The relation between   two observers in these two 
patches may be important as  one observer can move with acceleration for an other 
observer. In other words, in cosmological patch the black hole can move with 
acceleration leading to the appearence of two energy-fluxes: Hawking radiation and 
Unruh effect. The question is how to distinguish that.
Hence, finally the analysis gives almost the same situation for both 
patches. The small difference which appears in the evolution of black holes 
in two patches maybe understood taking into account also the above considerations.

The analysis of Nariai black hole evolution in modified Gauss-Bonnet 
gravity is also made. It is interesting that in this case the Nariai 
solution occurs if the model admits de Sitter solution but not necessary 
SdS solution where Gauss-Bonnet invariant is not constant. For specific 
realistic $f(G)$-gravity the very complicated numerical analysis is done.
It is demonstrated that for such theory the Nariai black hole remains to 
be stable. The study of Gauss-Bonnet gravity shows that our analysis maybe 
extended for other modified gravities: string-inspired theories, non-local 
gravities, gravity non-minimally coupled with matter, etc.
In particulary, the above (anti)-evaporation scenario has been recently 
investigated in $F(T)$ theory~\cite{DD}. However, as a rule the study of 
Nariai black hole evolution for other modified gravities turns out to be 
much more involved one than for $F(R)$-gravity.

\section*{Acknowledgments}
We would like to thank S. Zerbini and S. Nojiri for useful discussions and valuable suggestions. 
The work by SDO has been supported in part by MINECO (Spain), project FIS2010-15640, by AGAUR (Generalitat de Catalunya), contract 2009SGR-994 and by project 2.1839.2011 of MES (Russia).

\section{Appendix}
For the metric Ansatz (\ref{Nr3}), the non-zero components of Riemannian tensor read
\bea
R_{txtx}=e^{2\rho}(\rho''-\ddot{\rho})\,,\nonumber\\\nonumber\\
R_{t \theta t \theta}=-e^{-2\varphi}(\dot{\varphi}^2-\ddot{\varphi}+\dot{\varphi}\dot{\rho}+\varphi'\rho')\,,\nonumber\\\nonumber\\
R_{t \theta x \theta}=-e^{-2\varphi}(\dot{\varphi}\varphi'-\dot{\varphi'}+\dot{\varphi}\rho'+\dot{\rho}\varphi')\,,\nonumber\\\nonumber\\
R_{t \varphi t \varphi}=\sin^2\theta R_{t \theta t \theta}\,,\nonumber\\\nonumber\\
R_{t \varphi x \varphi}=-\sin^2\theta e^{-2\varphi}(-\dot{\varphi'}+\dot{\varphi}\varphi'+\dot{\varphi}\rho'+\dot{\rho}\varphi')\,,\nonumber\\\nonumber\\
R_{x \theta x \theta}=-e^{-2\varphi} (-\varphi''+\varphi'^2+\dot{\varphi}\dot{\rho}+\varphi'\rho')\,,\nonumber\\\nonumber\\
R_{\theta \varphi \theta \varphi}=-\sin^2\theta e^{-\varphi-\rho}(\dot{\varphi}^2 e^{-2\varphi} +e^{2\rho}-\varphi'^2e^{-2\varphi})\,.\nonumber
\eea

The Gauss-Bonnet invariant reads 
\bea%
G= -e^{-4\rho}\Big(-3\rho''^2+3\ddot{\rho}^2-3\ddot{\rho}\rho''+6\dot{\varphi}^2e^{2(\rho+\varphi)}
+28\varphi'\dot{\rho}\dot{\varphi}'+28\varphi'\dot{\varphi}\dot{\varphi}'-28\varphi''\dot{\rho}\dot{\varphi}\nonumber\\ \nonumber\\
+8\rho''e^{2(\rho+\varphi)}+28\dot{\varphi}'\dot{\varphi}\rho'-6e^{2(\rho+\varphi)}\varphi'^2-28\ddot{\varphi}\dot{\varphi}\dot{\rho}
-28\ddot{\varphi}\rho'\varphi'-28\varphi''\varphi'\rho'-8\ddot{\rho}e^{2(\rho+\varphi)}\nonumber\\ \nonumber\\
-16\varphi''\dot{\varphi}^2-16\ddot{\varphi}\varphi'^2-12\ddot{\varphi}\dot{\varphi}^2
+28\rho'\varphi'^3+16\varphi''\ddot{\varphi}-14\dot{\rho}^2\varphi'^2+28\varphi'^2\rho'^2
+28\varphi'^2\rho'^2\nonumber\\ \nonumber\\
-8\rho''\varphi'^2
-4\dot{\varphi}^2\varphi'^2-8\ddot{\rho}\dot{\varphi}^2+28\dot{\rho}\dot{\varphi}^3
+28\dot{\varphi}^2\dot{\rho}^2+8\ddot{\rho}\varphi'^2-12\varphi''\varphi'^2
+8\rho''\dot{\varphi}^2-14\dot{\varphi}^2\rho'^2\nonumber\\ \nonumber\\
+3e^{4(\rho+\varphi)}+28\varphi'\rho'\dot{\varphi}\dot{\rho}+6\ddot{\varphi}^2-14\dot{\varphi}'^2
+6\varphi''^2+9\dot{\varphi}^2+9e^{4\rho}\varphi'^2\Big)\,.\nonumber
\eea%
The $(0,0)$ component of $f(G)$-field equations (\ref{GB-EOM}) results to be
\bea
&&-\ddot{\rho}+2\ddot{\varphi}+\rho''-2\dot{\varphi}^2-2\dot{\rho}\dot{\varphi}
-2\rho'\varphi'+\frac{1}{2}Re^{2\rho}=-\frac{1}{2}fe^{2\rho}
-2FR(-\ddot{\rho}+2\ddot{\varphi}+\rho''-2\dot{\varphi}^2-2\dot{\rho}\dot{\varphi}-2\rho'\varphi')
\nonumber\\ \nonumber\\
&&+4Fe^{-2\rho}\Big[(2\dot{\varphi}'-2\varphi'\dot{\varphi}-2\rho'\dot{\varphi}-2\dot{\rho}\varphi')^2
-(-\ddot{\rho}+2\ddot{\varphi}+\rho''-2\dot{\varphi}^2-2\dot{\rho}\dot{\varphi}-2\rho'\varphi')^2\Big]
\nonumber\\ \nonumber\\
&&-2Fe^{-2\rho}\Big[-(\rho''-\ddot{\rho})^2-2(\dot{\varphi}^2-\ddot{\varphi}+\dot{\varphi}\dot{\rho}
+\varphi'\rho')^2
+2(-\dot{\varphi}'+\dot{\varphi}\varphi'+\dot{\varphi}\rho'+\varphi'\dot{\rho})^2\Big]
\nonumber\\ \nonumber\\
&&-4F\Big[-e^{-2\rho}(\rho''-\ddot{\rho})(-\rho''+\ddot{\rho}+2\varphi''-2\varphi'^2-2\dot{\rho}\dot{\varphi}
-2\rho\varphi')+2e^{2\varphi}(\dot{\varphi}^2-\ddot{\varphi}+\dot{\varphi}\dot{\rho}+\rho'\varphi')
\nonumber\\ \nonumber\\
&&(1+e^{-2\rho-2\varphi}(-\ddot{\varphi}+\varphi''+2\dot{\varphi}^2-2\varphi'^2))\Big]
+2R(\ddot{F}-\dot{\rho}\dot{F}-\rho' F')+2Re^{2\varphi}((e^{-2\varphi}F')_{x}-(e^{-2\varphi}\dot{F})_t)\nonumber\\ \nonumber\\
&&-8e^{-2\rho} \Big[(-\rho''+\ddot{\rho}+2\varphi''-2\varphi'^2-2\dot{\varphi}\dot{\rho}-2\varphi'\rho')
(\dot{F}'-\rho'\dot{F}-\dot{\rho}F')-(-\ddot{\rho}+2\ddot{\varphi}+\rho''-2\dot{\varphi}^2
-2\dot{\varphi}\dot{\rho}\nonumber\\ \nonumber\\
&&-2\rho'\varphi')(\ddot{F}-\dot{\rho}\dot{F}-\rho'F')\Big]+
4e^{2\varphi-2\rho}(-\ddot{\rho}+2\ddot{\varphi}+\rho''-2\dot{\varphi}^2-2\dot{\varphi}\dot{\rho}
-2\rho'\varphi')((e^{-2\varphi}F')_{x}-(e^{-2\varphi}\dot{F})_t)\nonumber\\ \nonumber\\
&&-4e^{\rho}\Big(e^{-3\rho}(\ddot{F}-\dot{\rho}\dot{F}-\rho'F')(-\ddot{\rho}
+2\ddot{\varphi}+\rho''-2\dot{\varphi}^2-2\dot{\varphi}\dot{\rho}-2\rho'\varphi')
-2e^{-3\rho}(\dot{F}'-\rho'\dot{F}-\dot{\rho}F')(2\dot{\varphi}'-2\varphi'\dot{\varphi}
\nonumber\\ \nonumber\\
&&-2\rho'\dot{\varphi}-2\dot{\rho}\varphi')-2e^{3\varphi}(\varphi'F'-\dot{\varphi}\dot{F})
(1+e^{-2\rho-2\varphi}(-\ddot{\varphi}+\varphi''+2\dot{\varphi}^2-2\varphi'^2)\Big)
\nonumber\\ \nonumber\\
&&+4e^{-\rho}\Big[e^{-\rho}(\rho''-\ddot{\rho})(F''-\dot{\rho}\dot{F}-\rho' F')+2e^{\varphi}(\varphi'F'-\dot{\varphi}\dot{F})(\dot{\varphi}^2-\ddot{\varphi}+\dot{\rho}\dot{\varphi}+\varphi'\rho')\Big]=0
\label{tt}.\nonumber
\eea
\par
For $(1,1)$ component we have
\bea
&&-\rho''+\ddot{\rho}+2\varphi''-2\varphi'^2-2\dot{\varphi}\dot{\rho}-2\rho'\varphi'-\frac{1}{2}e^{2\rho}R
=\frac{1}{2}fe^{2\rho}-2FR(-\rho''+\ddot{\rho}+2\varphi''-2\varphi'^2-2\varphi'\rho'-2\dot{\varphi}\dot{\rho})
\nonumber\\ \nonumber\\ &&
+4Fe^{-2\rho}\Big[(-\rho''+\ddot{\rho}+2\varphi''-2\varphi'^2
-2\dot{\varphi}\dot{\rho}-2\varphi'\rho')^2-(2\dot{\varphi'}-2\varphi'\dot{\varphi}
-2\rho'\dot{\varphi}-2\dot{\rho}\varphi')^2\Big]
\nonumber\\ \nonumber\\ &&
-2Fe^{-2\rho}\Big[-(\rho''-\ddot{\rho})^2
-2(\dot{\varphi}\dot{\varphi}-\dot{\varphi}'+\dot{\varphi}\rho'+\dot{\rho}\varphi')^2
+(-\varphi''+\varphi'^2+\dot{\varphi}\dot{\rho}+\varphi'\rho')^2\Big]
\nonumber\\ \nonumber\\ &&
-4F\Big[e^{-2\rho}(\rho''-\ddot{\rho})(-\ddot{\rho}+2\varphi''+\rho''-2\dot{\varphi}^2
-2\dot{\varphi}\dot{\rho}-2\rho'\varphi')+e^{2\varphi}(1+e^{-2\varphi-2\rho}(-\ddot{\varphi}
\nonumber\\ \nonumber\\ &&
+\varphi''+2\dot{\varphi}^2-2\varphi'^2))(-\varphi''+\varphi'^2+\dot{\varphi}\dot{\rho}+\rho'\varphi')\Big]
\nonumber\\ \nonumber\\ &&
+2R(F''-\dot{\rho}\dot{F}-\rho' F')-2Re^{2\varphi}((e^{-2\varphi}F')_{x}-(e^{-2\varphi}\dot{F})_{\tau})
-8e^{-2\rho}\Big[(-\rho''+\ddot{\rho}+2\varphi''-2\varphi'^2-2\dot{\varphi}\dot{\rho}-2\rho'\varphi')
\nonumber\\ \nonumber\\ &&
(F''-\dot{\rho}\dot{F}-\rho' F')-(2\dot{\varphi}'-2\varphi'\dot{\varphi}-2\rho'\dot{\varphi}-2\dot{\rho}\varphi')
(F''-\rho'\dot{F}-\dot{\rho}F')\Big]+4e^{2\varphi-2\rho}(-\rho''+\ddot{\rho}+2\varphi''-2\varphi'^2
\nonumber\\ \nonumber\\ &&
-2\dot{\varphi}\dot{\rho}-2\rho'\varphi')((e^{-2\varphi}F')_{x}-(e^{-2\varphi}\dot{F})_{\tau})
+4e^{\rho}\Big[\Big(e^{-3\rho}(\ddot{F}-\dot{\rho}\dot{F}-\rho'F')(-\ddot{\rho}+2\ddot{\varphi}
+\rho''-2\dot{\varphi}^2
\nonumber\\ \nonumber\\ &&
-2\dot{\varphi}\dot{\rho}-2\rho'\varphi')-2e^{-3\rho}(\dot{F}'-\rho'\dot{F}
-\dot{\rho}F')(2\dot{\varphi}'-2\varphi'\dot{\varphi}-2\rho'\dot{\varphi}-2\dot{\rho}\varphi')
\nonumber\\ \nonumber\\ &&
-2e^{3\varphi}(\varphi'F'-\dot{\varphi}\dot{F})(1+e^{-2\rho-2\varphi}(-\ddot{\varphi}+\varphi''
+2\dot{\varphi}^2-2\varphi'^2)\Big)\Big]
\nonumber\\ \nonumber\\ &&
-4e^{-\rho}\Big[e^{-\rho}(\rho''-\ddot{\rho})(\ddot{F}-\dot{\rho}\dot{F}-\rho' F')-e^{\varphi}(\varphi'F'-\dot{\varphi}\dot{F})(\varphi'^2-\varphi''+\dot{\rho}\dot{\varphi}+\varphi'\rho')\Big]=0
\label{xx}.\nonumber
\eea
For $(0,1)$($=(1,0)$) component one obtains
\bea
&&2\dot{\varphi}'-2\varphi'\dot{\varphi}-2\rho'\dot{\varphi}-2\dot{\rho}\varphi'=-2FR(2\dot{\varphi}'
-2\varphi'\dot{\varphi}-2\rho'\dot{\varphi}-2\dot{\rho}\varphi')
\nonumber\\ \nonumber\\ &&
+4Fe^{-2\rho}\Big[(2\dot{\varphi}'-2\varphi'\dot{\varphi}-2\rho'\dot{\varphi}
-2\dot{\rho}\varphi')(-2\rho''+2\ddot{\rho}+
2\varphi''-2\varphi'^2-2\ddot{\varphi}+2\dot{\varphi}^2)\Big]
\nonumber\\ \nonumber\\ &&
-4Fe^{-2\rho}\Big[\dot{\varphi}^2-\ddot{\varphi}+\dot{\varphi}\dot{\rho}+\rho'\varphi'\Big]
\Big[-\dot{\varphi}'+\dot{\varphi}\varphi'
+\dot{\varphi}\rho'+\dot{\rho}\varphi'\Big]
-4F\Big(-2e^{-2\rho}(\rho''-\ddot{\rho})(2\dot{\varphi}'-2\varphi'\dot{\varphi}
\nonumber\\ \nonumber\\ &&
-2\rho'\dot{\varphi}-2\dot{\rho}\varphi')
+2e^{2\varphi}(\varphi\dot{\varphi}-\dot{\varphi}'+\dot{\varphi}\rho'+\varphi'\dot{\rho})
(1+e^{-2\varphi-2\rho}(-\ddot{\varphi}+\varphi''+2\dot{\varphi}^2-2\varphi'^2))\Big)
\nonumber\\ \nonumber\\ &&
+2R(\dot{F}'-\dot{\rho}F'-\rho'\dot{F})-4e^{-2\rho}\Big[(2\dot{\varphi}'-2\varphi'\dot{\varphi}
-2\rho'\dot{\varphi}-2\dot{\rho}\varphi')(F''-\dot{\rho}\dot{F}-\rho'  F')
\nonumber\\ \nonumber\\ &&
-(\ddot{\rho}+2\ddot{\varphi}+\rho''-2\dot{\varphi}^2-2\dot{\rho}\dot{\varphi}-2\rho'\varphi')
(\dot{F}'-\dot{\rho}F'-\rho'\dot{F})\Big]
-4e^{-2\rho}\Big[(\dot{F}'-\dot{\rho}F'-\rho'\dot{F})(-\rho''+\ddot{\rho}+2\varphi''
\nonumber\\ \nonumber\\ &&
-2\varphi'^2-2\dot{\rho}\dot{\varphi}-2\rho'\varphi')-(\ddot{F}-\dot{\rho}\dot{F}-\rho' F')(2\dot{\varphi}'
-2\varphi'\dot{\varphi}-2\dot{\varphi}\rho'-2\dot{\rho}\varphi')\Big]
+4e^{2\varphi-2\rho}(2\dot{\varphi}'-2\varphi\dot{\varphi}-2\rho'\dot{\varphi}
\nonumber\\ \nonumber\\ &&
-2\dot{\rho}\varphi')
((e^{-2\varphi}F')_x-(e^{-2\varphi}\dot{F})_\tau)-4\Big(e^{-2\rho}(\rho''-\ddot{\rho})(\dot{F}'-\dot{\rho}F'
-\rho'\dot{F})-2e^{\varphi-\rho}(\dot{\varphi}\varphi'-\dot{\varphi}'+\dot{\varphi}\rho'+\varphi'\dot{\rho})\nonumber\\\nonumber\\&&
(\varphi'F'-\dot{\varphi}\dot{F})\Big)=0\,.
\label{xt}\nonumber
\eea
Finally, the ($2,2$) (=($3,3$)) component reads
\bea
&&1+e^{-2(\rho+\varphi)}(-\ddot{\varphi}+\varphi''+2\dot{\varphi}^2-2\varphi'^2)
-\frac{1}{2}Re^{-2\varphi}=\frac{1}{2}fe^{-2\varphi}-2FR\Big[1+e^{-2(\rho+\varphi)}
\nonumber\\ \nonumber\\ &&
(-\ddot{\varphi}+\varphi''+2\dot{\varphi}^2-2\varphi'^2)\Big]+4Fe^{-2\varphi}\Big[1+e^{-2(\rho+\varphi)}
(-\ddot{\varphi}+\varphi''+2\dot{\varphi}^2-2\varphi'^2)\Big]^2-2F\Big[-e^{-2\varphi
-4\rho}(\dot{\varphi}^2
\nonumber\\ \nonumber\\ &&
-\ddot{\varphi}+\dot{\varphi}\dot{\rho}+\varphi'\rho')^2
-e^{-2\varphi-4\rho}(\dot{\varphi}\varphi'-\dot{\varphi'}+\dot{\varphi}\rho'
+\varphi'\dot{\rho})^2+e^{-2\varphi-4\rho}(-\varphi''+\varphi'^2+\dot{\varphi}\dot{\rho}
\nonumber\\ \nonumber\\ &&
+\varphi'\rho')^2-e^{4\varphi-2\rho}(\dot{\phi}^2e^{-2\varphi}+e^{2\rho}-\varphi'^2e^{-2\rho})^2\Big]
-4F\Big[e^{-2\varphi-4\rho}(\dot{\varphi}^2-\ddot{\varphi}+\dot{\varphi}\dot{\rho}+\varphi'\rho')
\nonumber\\ \nonumber\\ &&
(-\ddot{\rho}+2\ddot{\varphi}+\rho''-2\dot{\varphi}^2-2\dot{\rho}\dot{\varphi}-2\varphi'\rho')
+e^{-2\varphi-4\rho}(2\dot{\varphi'}-2\varphi'\dot{\varphi}-2\rho'\dot{\varphi}-2\dot{\rho}\varphi')
\nonumber\\ \nonumber\\ &&
(\dot{\varphi}\varphi'-\dot{\varphi}'+\dot{\varphi}\rho'+\dot{\rho}\varphi')+e^{-2\varphi-4\rho}
(-\varphi''+\varphi'^2+\dot{\varphi}\dot{\rho}+\varphi'\rho')(-\rho''+\ddot{\varphi}+2\varphi''
\nonumber\\ \nonumber\\ &&
-2\varphi'^2-2\dot{\varphi}\dot{\rho}-2\varphi'\rho')+e^{3\varphi-\rho}(\dot{\varphi}^2e^{-2\rho}
+e^{2\rho}-\varphi'^2e^{-2\varphi})(1+e^{-2(\rho+\varphi)}(-\ddot{\varphi}+\varphi''+2\dot{\varphi}^2
\nonumber\\ \nonumber\\ &&
-2\varphi'^2))\Big]
-2Re^{-\rho-\varphi}(\varphi'F'-\dot{\varphi}\dot{F})-2Re^{-2\rho}((e^{-2\varphi}F')_x-(e^{-2\varphi}\dot{F})_\tau)
-8Fe^{\varphi-\rho}(\dot{\varphi}\dot{F}-\varphi' F')
\nonumber\\ \nonumber\\ &&
(1+e^{-2(\rho+\varphi)}(-\ddot{\varphi}+\varphi''+2\dot{\varphi}^2-2\varphi'^2))
+4e^{-2\rho+2\varphi}(1+e^{-2(\rho+\varphi)}(-\ddot{\varphi}+\varphi''+2\dot{\varphi}^2
\nonumber\\ \nonumber\\ &&
-2\varphi'^2))((e^{-2\varphi}F')_x-(e^{-2\varphi}\dot{F})_\tau)
+4e^{-4\rho-2\varphi}\Big[(\ddot{F}-\dot{F}\dot{\rho}-\rho'  F')(-\ddot{\rho}+2\ddot{\varphi}+\rho''-2\dot{\varphi}^2-2\dot{\rho}\dot{\rho}-2\rho \varphi')
\nonumber\\ \nonumber\\ &&
-2(\dot{F}'-\rho'\dot{F}-\dot{\rho}F')(2\dot{\varphi}'-2\varphi'\dot{\varphi}
-2\rho'\dot{\varphi}-2\dot{\rho}\varphi')+(F''-\dot{\rho}\dot{F}-\rho' F')
\nonumber\\ \nonumber\\ &&
(-\rho''+\ddot{\rho}+2\varphi''-2\varphi'^2-2\dot{\varphi}\dot{\rho}-2\varphi'\rho')\Big]
-4\Big[e^{-2\varphi-4\rho}(\dot{\varphi}^2-\ddot{\varphi}+\dot{\varphi}\dot{\rho}+\varphi'\rho')
\nonumber\\ \nonumber\\ &&
(\ddot{F}-\dot{\rho}\dot{F}-\rho' F')-e^{-2\varphi-4\rho}(\dot{\varphi}\varphi'-\dot{\varphi}'+\dot{\varphi}\rho'+\dot{\rho}\varphi')(F''-\dot{\rho}F'- \rho'\dot{F})
\nonumber\\ \nonumber\\ &&
+e^{-2\varphi-4\rho}(-\varphi''+\varphi'^2+\dot{\varphi}\dot{\rho}+\varphi'\rho')(F''-\dot{\rho}\dot{F}-\rho' F')+e^{3\varphi-\rho}(\dot{\varphi}^2e^{-2\varphi}+e^{2\rho}-\varphi'^2 e^{-2\varphi})(\varphi' F'-\dot{\varphi}\dot{F})\Big]=0\,.\nonumber\\\label{thetatheta}\nonumber
\eea

\end{document}